\documentclass[lettersize,journal]{IEEEtran}
\usepackage{amsmath,amssymb,amsfonts}
\usepackage{algorithmic}
\usepackage{array}
\usepackage[caption=false,font=normalsize,labelfont=sf,textfont=sf]{subfig}
\usepackage{textcomp}
\usepackage{stfloats}
\usepackage{url}
\usepackage{verbatim}
\usepackage{graphicx}
\usepackage{cite}

\usepackage{multirow}
\usepackage{textcomp}
\usepackage{xcolor}
\usepackage{tabularx}
\usepackage{arydshln}
\usepackage{booktabs}
\usepackage{hyperref}

\def\BibTeX{{\rm B\kern-.05em{\sc i\kern-.025em b}\kern-.08em
    T\kern-.1667em\lower.7ex\hbox{E}\kern-.125emX}}

\begin{document}

\title{BYOL for Audio: Exploring Pre-trained General-purpose Audio Representations}

\author{Daisuke~Niizumi, Daiki~Takeuchi, Yasunori~Ohishi,~\IEEEmembership{Member,~IEEE,} Noboru~Harada,~\IEEEmembership{Senior~Member,~IEEE,} and~Kunio~Kashino,~\IEEEmembership{Senior~Member,~IEEE}
\thanks{Manuscript received April 19, 2021; revised August 16, 2021.}%
\thanks{The authors are with Communication Science  Laboratories, Nippon Telegraph and Telephone Corporation,  Atsugi 243-0198, Japan (e-mail: daisuke.niizumi.dt@hco.ntt.co.jp; daiki.takeuchi.ux@hco.ntt.co.jp; yasunori.ooishi.uk@hco.ntt.co.jp; noboru.harada.pv@hco.ntt.co.jp; kunio.kashino.me@hco.ntt.co.jp)}
}

\markboth{PREPRINT}%
{Niizumi \MakeLowercase{\textit{et al.}}: BYOL for Audio: Exploring Pre-trained General-purpose Audio Representations}

\IEEEpubid{PREPRINT}


\maketitle

\begin{abstract} 
Pre-trained models are essential as feature extractors in modern machine learning systems in various domains. In this study, we hypothesize that representations effective for general audio tasks should provide multiple aspects of robust features of the input sound.
For recognizing sounds regardless of perturbations such as varying pitch or timbre, features should be robust to these perturbations.
For serving the diverse needs of tasks such as recognition of emotions or music genres, representations should provide multiple aspects of information, such as local and global features.
To implement our principle, we propose a self-supervised learning method: Bootstrap Your Own Latent (BYOL) for Audio (BYOL-A, pronounced ”viola”). BYOL-A pre-trains representations of the input sound invariant to audio data augmentations, which makes the learned representations robust to the perturbations of sounds. Whereas the BYOL-A encoder combines local and global features and calculates their statistics to make the representation provide multi-aspect information.
As a result, the learned representations should provide robust and multi-aspect information to serve various needs of diverse tasks.
We evaluated the general audio task performance of BYOL-A compared to previous state-of-the-art methods, and BYOL-A demonstrated generalizability with the best average result of 72.4\% and the best VoxCeleb1 result of 57.6\%.
Extensive ablation experiments revealed that the BYOL-A encoder architecture contributes to most performance, and the final critical portion resorts to the BYOL framework and BYOL-A augmentations. 
Our code is available online for future studies.
\end{abstract}

\section{Introduction}
\IEEEPARstart{P}{re}-trained models play a vital role as feature extractors in various domains, e.g., BERT\cite{bert} in the natural language processing domain and ImageNet pre-trained models\cite{vgg,resnet,resnext} in the image domain.
In the audio domain, pre-trained models (e.g., VGGish\cite{hershey2017cnn}) have enabled recent advances in applications such as heart sound classification\cite{koike2020heartsound}, Alzheimer's disease detection\cite{balagopalan2021alzheimer}, conservation monitoring\cite{sethi2020soundscapes}, audio captioning\cite{koizumi2020tfmcaptioning}, audio retrieval\cite{oncescu2021audioretrieval}, and so forth.

\begin{figure}[htb]
  \centering
  \includegraphics[bb=0 0 460 320, width=\columnwidth]{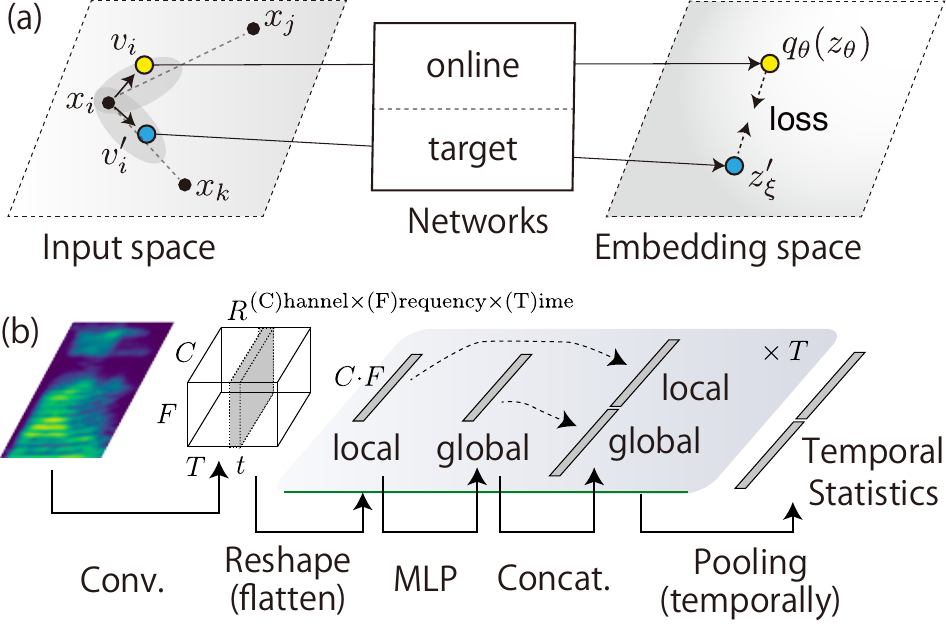} 
      \caption{(a) BYOL\cite{grill2020byol} for Audio learning scenario.
      An input audio $x_i$ branches into two directions, or views, $v_i$ and $v_i'$, by mixing audio $x_j$ and $x_k$ and making pitch/time/amplitude random.
      $v_i$, $v_i'$ are projected through networks, and then loss is minimized on the projected $z'_\xi$ and predicted $q_\theta(z_\theta)$. BYOL-A learns representations invariant to the difference between $v_i$ and $v_i'$. Find more details in Section \ref{sec:byol}.
      (b) BYOL-A feature calculation: The input becomes feature maps extracted by Conv. For each time frame, the feature map is flattened as local features by Reshape, turned into global features by MLP, then both features concatenate into mixed features. Finally, T frame features are temporally pooled into mean+max statistics, achieving a multi-aspect robust feature.}
  \label{fig:scenario}
  \vspace{-10pt}
\end{figure}

Various audio pre-trained models have been proposed for supervised learning\cite{kong2020panns,gong2021psla,guzhov2021esresnext,gong2021ast} or unsupervised learning\cite{favory2020coala,cramer2019openl3,wang2021multimodal,shor2020trill,saeed2020cola,fonseca2020uclser20} methods, and they have been evaluated on target tasks such as sound event recognition (SER)\cite{gemmeke2017audioset,piczak2015esc50,salamon2014urbansound,fonseca2020fsd50k}, non-semantic speech (NOSS)\cite{shor2020trill} (e.g., speech command recognition\cite{speechcommandsv2}, speaker identification\cite{voxceleb}), and music tasks (e.g., music genre\cite{gt2013gtzan} and instrument\cite{nsynth2017} classification).
However, while the methods claim the state-of-the-art, it is unclear which method generalizes better because their benchmarks are not compatible.

Our goal is to explore a way to achieve a versatile audio representation that works effectively for various tasks as it is, off-the-shelf, without an extra effort such as fine-tuning.
The applicability of a model increases if it can be used as a frozen feature extractor because the effort for fine-tuning is not negligible, such as a careful learning rate tuning not to break pre-trained valuable features.
Thus, if a representation is versatile enough as it is, it is an ultimate goal.

\IEEEpubidadjcol

However, various task settings are both common and conflicting, making a single one-fits-all representation difficult.
For example, we recognize words regardless of who speaks; conversely, we identify speakers while ignoring speech content words.
In contrast to these conflicts, we commonly ignore slight differences such as pitch, duration, or timbre when listening speech.
These suggest that while multiple information may serve conflicting needs, ignoring slight differences may serve the common needs.

For serving different needs, multiple features available from different layers of a single model and statistics of these features potentially be helpful.
The former study \cite{alain2016understanding} showed that early layers, local features on CNNs, contain relatively "general" filters, whereas deeper layers, global features, are specific to the pre-training. The studies \cite{chenhui2019multilayers,weisen2021seen,yue2018elastic} even utilized fusing the multilayer features.
In addition, the global pooling for summarizing variable-length audio features provides multiple options, such as temporal average or max pooling\cite{favory2020coala}\cite{Kumar2021DoSE}\cite{fonseca2021improving}, or even combining them\cite{kong2020panns}.

For serving common needs, we want a representation that ignores slight differences in sounds, and self-supervised learning (SSL) frameworks for the image domain can be a good choice. These methods learn representations invariant to augmentations, and we can use audio data augmentations to make a difference in sounds.
Typical choices are contrastive learning methods such as SimCLR\cite{chen20simclr} or MoCo\cite{he2020momentum}, which learn representations through discriminating augmented positive pairs from augmented negative pairs in an input batch.
However, we think Bootstrap Your Own Latent (BYOL)\cite{grill2020byol} can be a better choice because it learns representations invariant to changes created by augmentations, achieving what we seek directly.

To implement our principle, we combine the aforementioned options to achieve a versatile general-purpose audio representation.
For encoding such representation, our network takes multilayer features and calculates statistics for accommodating multi-aspect information, which serves different needs of tasks.
For pre-training, our BYOL variant framework with audio data augmentations learns a representation robust to the perturbations created by the augmentations, which serves common needs of tasks.
As a result, the learned representations should provide multi-aspect robust features of the input sounds and serve various needs of diverse tasks.


The following summarizes our contributions:

\begin{itemize}
    \item We propose Bootstrap Your Own Latent (BYOL) for Audio (BYOL-A, pronounced "viola"). BYOL-A learns representations robust to the perturbations of sounds, and its encoder combines statistics of local and global features to provide multiple aspects of information.
    \item We make a new benchmark that evaluates generalizability on diverse audio tasks.
    \item We demonstrate the generalizability and effectiveness of our method using the benchmark while comparing ours with eleven representations extracted from the conventional state-of-the-art models.
    \item We conduct intensive ablation studies to clarify the contributions of the BYOL-A framework, augmentations, and encoder network architecture.
    \item We make our code available online\footnote{\url{https://github.com/nttcslab/byol-a}} for reproducibility and to foster progress in future studies.
\end{itemize}

Fig. \ref{fig:scenario} describes the BYOL-A representation learning scenario and feature calculation on the encoder.
BYOL-A pre-trains the encoder to transform the input into representations robust to data augmentations, and its encoder combines statistics of local and global features to provide multiple aspects of information. As a result, representations should become robust and multi-aspect for serving various needs of audio task settings.

\section{Related Work}

\subsection{Relationship with our previous work}
We introduced BYOL-A in our previous work\cite{niizumi2021byol-a}. In this section, we clarify the relationship between the previous and present work.

Our previous work proposed BYOL-A, which extends BYOL to work with audio data augmentations designed to learn the audio representations of specific targets, namely foreground acoustic event sound and the sound texture details.
Though it showed the state-of-the-art performance by learning representations for these targets, the comparisons were limited only among unsupervised learning methods, and we did not discuss encoder architecture improvements.

In this paper, we redefined our hypothesis to explore pre-trained general-purpose audio representations with a broader research scope. To achieve this new goal, we extended BYOL-A to learn representations invariant to the perturbations of sound. We also extended the BYOL-A encoder architecture to provide multiple aspects of learned features.
Ablation studies in Sections \ref{sec:exp-abl-byola}, \ref{sec:exp-abl-byola-arch}, and \ref{sec:exp-abl-gp} clarify the improvements from the previous one proposed in \cite{niizumi2021byol-a}.

Detailed differences are listed as follows:
\begin{itemize}
\item We redefine our hypothesis for effective general-purpose audio representations.
\item We reinterpret BYOL-A to learn representations invariant to the perturbations of sounds made by data augmentations rather than learn representations of specific sounds.
\item We refine data augmentations for improving performance.
\item We improve the encoder architecture to combine multiple aspects of information.
\item We evaluate our proposals with a wide variety of popular models and tasks under a unified benchmark.
\item We conduct intensive ablation studies to analyze contributions of the BYOL-A framework, augmentations, and encoder network architecture.
\end{itemize}

\subsection{Audio pre-training methods}\label{sec:relworks-prev-models}
First, we overview previous pre-training methods and differentiate ours from them.
Supervised learning methods learn representations to discriminate labels, relying on labels assigning samples to a predefined class\cite{hershey2017cnn,kong2020panns,gong2021psla,guzhov2021esresnext,gong2021ast}.
Self-supervised learning (SSL) methods using masked prediction predominant in the speech domain learn to predict or reconstruct masked portions of the input, relying on the input masking\cite{Liu2020Mockingjay,baevski2020wav2vec2,Hsu2021HuBERT}.
SSL methods using contrastive learning learn to discriminate instances among batch samples, relying on comparison in a large batch\cite{saeed2020cola,fonseca2020uclser20,spijkervet2021contrastive,shor2020trill}.
Cross-modal SSL methods also learn to discriminate correspondence across modalities, relying on the cooccurrence of the pair of modalities\cite{cramer2019openl3,wang2021multimodal,favory2020coala}.
We adopt BYOL and learn representations invariant to input changes, relying on the changes of input audio created by audio data augmentations.

Many supervised learning models pre-trained on large-scale datasets have been proposed mainly for SER tasks.
VGGish\cite{hershey2017cnn} pre-trained on YouTube-8M\cite{youtube8m} is used as a feature extractor in various application studies\cite{sethi2020soundscapes}\cite{koizumi2020tfmcaptioning}\cite{oncescu2021audioretrieval}\cite{kim2019audiocaps}\cite{tolkova2021parsing}.
Pre-trained Audio Neural Networks (PANNs)\cite{kong2020panns} models pre-trained on AudioSet\cite{gemmeke2017audioset} showed state-of-the-art results, and they have been used in application studies \cite{koike2020heartsound}\cite{tolkova2021parsing}.
Pre-training on both ImageNet and AudioSet shows advantages: PSLA\cite{gong2021psla}, ESResNe(X)t-fbsp\cite{guzhov2021esresnext}, and Audio Spectrogram Transformer (AST)\cite{gong2021ast} produced state-of-the-art results on SER.

For unsupervised learning, self-supervised learning (SSL) methods have been proposed including COLA\cite{saeed2020cola} for general-purpose audio representations, Fonseca et al.\cite{fonseca2020uclser20} for SER, and CLMR\cite{spijkervet2021contrastive} for music tasks.
SSL models are specifically proposed for speech representations, such as TRILL\cite{shor2020trill}, PACE+\cite{ravanelli2020paceplus}, Mockingjay\cite{Liu2020Mockingjay}, Wav2Vec 2.0\cite{baevski2020wav2vec2}, and HuBERT\cite{Hsu2021HuBERT}.
In particular, non-speech recognition application studies such as \cite{balagopalan2021alzheimer} use Wav2Vec 2.0 as a feature extractor.

Cross-modal/multi-modal pre-training methods have been proposed. OpenL3\cite{cramer2019openl3} was pre-trained by using audio-visual correspondence as training signal. The method by Wang et al. \cite{wang2021multimodal} was pre-trained by using correspondence between video, spectrograms, and raw waveforms to learn general-purpose audio representations.
COALA\cite{favory2020coala} was pre-trained by aligning the learned latent representations of audio and associated tags and evaluated on SER and music tasks. For speech tasks, a concurrent work SLAM\cite{bapna2021slam} learns speech and language modeling jointly in a multi-task fashion.

These methods showed effectiveness, but their evaluation settings are not compatible with each other, making comparison difficult for future applications to pick a suitable representation.

In concurrent works, BigSSL\cite{zhang2021bigssl}, data2vec\cite{baevski2022data2vec}, and the method by Wang et al. \cite{wang2022universal} exhibit remarkable performance on various tasks, while SERAB\cite{scheidwasserclow2021serab} adopts our previous BYOL-A\cite{niizumi2021byol-a} on speech emotion recognition tasks.
The data2vec combines masked prediction and learning the latent target representations, similar to BYOL.
WavLM\cite{Chen2021WavLM} learns representations using masked speech prediction and denoising mixed utterances, and the learning from denoising is similar to ours. BYOL-A learns a representation of a sound invariant to the mixed background sound.

\subsection{Benchmarks for pre-trained models}
To assess the generalizability and re-usability of pre-trained models across a wide range of tasks, we need a benchmark such as SUPERB\cite{yang2021superb}.
Like the standard linear evaluation protocol\cite{oord2018cpc, chen20simclr}, SUPERB trains lightweight heads on top of the frozen pre-trained model\footnote{While SUPERB follows a fashion similar to the standard linear evaluation, it collects multiple features from different model layers and weighted-sum them as the representations to evaluate, which makes some task results, such as VoxCeleb1, better than the trend in our results.}.
While it supports a broad range of tasks, these tasks are limited to the speech domain; no established benchmark exists for non-speech audio tasks.
Therefore we build a benchmark for evaluating general-purpose audio representations in this study.

Concurrent works on benchmarks share a setting similar to that in this study to evaluate the generalizability of \textit{frozen} pre-trained models across different architectures, pre-training frameworks, and datasets. HARES\cite{wang2022universal} trains a linear layer head, the same as we do, while HEAR\cite{turian2022hear} uses a shallow MLP downstream classifier. SERAB\cite{scheidwasserclow2021serab} also evaluates \textit{frozen} models in various speech emotion recognition tasks.

\subsection{Global pooling design choices}\label{sec:global-pooling-designs}
Previous studies have taken various global pooling approaches:
VGGish\cite{hershey2017cnn}, OpenL3\cite{cramer2019openl3}, and COALA\cite{favory2020coala} flatten frequency bins, time frames, and channels into a single embedding vector. 
COLA\cite{saeed2020cola}, TRILL\cite{shor2020trill}, ESResNe(X)t-fbsp\cite{guzhov2021esresnext}, and Fonseca et al.\cite{fonseca2020uclser20} use the global average or max pooling commonly used in image CNNs. 

Other approaches average frequency first and then summarize time.
PANNs'\cite{kong2020panns} CNN14 model first averages the frequency\footnote{\scriptsize \url{https://github.com/qiuqiangkong/audioset_tagging_cnn/blob/master/pytorch/models.py\#L215}}. Then, it applies a temporal pooling operation, which we refer to as \textit{temporal mean+max pooling} hereafter, that calculates a sum of both temporal mean- and max-pooling as the resulting output.
Fonseca et al. \cite{fonseca2021improving} also used frequency average pooling first and then applied temporal max pooling.
Ford et al.\cite{ford2019deepresidual} proposed adding an attention module to summarize the output of frequency average pooling.

All these approaches apply flattening, max, or averaging operations, which can impair the information needed for downstream tasks. For example, averaging along frequency hides information about frequency patterns that can be crucial to tasks such as speaker age estimation.

\subsection{Bootstrap Your Own Latent (BYOL)}\label{sec:byol}

BYOL\cite{grill2020byol} is a self-supervised learning algorithm that learns image representations invariant to data augmentations.
Although contrastive learning methods such as SimCLR\cite{chen20simclr} and MoCo\cite{he2020momentum} also learn representations invariant to data augmentations as BYOL does, we believe that BYOL is appropriate for our purpose because it learns representations from a single input, whereas contrastive learning methods learn by comparison among input batch samples.

\begin{figure*}[htb!]
  \centering
  \includegraphics[width=1.0\textwidth]{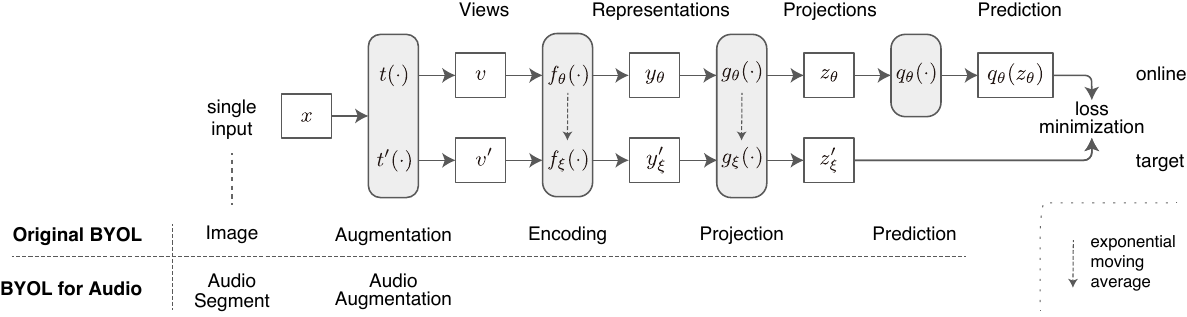} 
  \vspace{-3pt}
  \caption{Original BYOL and BYOL for Audio system overview.}
  \label{fig:byola-system}
  \vspace{-7pt}
\end{figure*}

As shown in Fig.~\ref{fig:byola-system}, BYOL consists of two neural networks, referred to as online and target networks.
The online network is defined by a set of weights $\theta$, and the target network has the same architecture as the online network but uses a different set of weights, $\xi$.
First, BYOL produces two augmented views, $v\triangleq t(x)$ and $v'\triangleq t'(x)$, from an image $x$ by applying respectively image augmentations $t\sim \mathcal{T}$ and $t'\sim \mathcal{T}'$, where $\mathcal{T}$ and $\mathcal{T}'$ denote the two distributions of the image augmentations.
Then, the online network outputs a representation $y_{\theta}$, a projection $z_{\theta}$, and a prediction $q_{\theta}(z_{\theta})$  from the first view $v$.
On the other hand, the target network outputs $y'_{\xi}$ and the target projection $z'_{\xi}$ from the second view~$v'$.
Finally, the following mean squared error between the L2-normalized predictions $\overline{q_\theta}(z_\theta)$ and target projections $\overline{z}'_\xi$ is calculated:

\vspace{-10pt}
\begin{equation}
L_{\theta,\xi} \triangleq ||\overline{q_\theta}(z_\theta) - \overline{z}'_\xi||^2_2 = 2 - 2 \cdot \frac{\langle q_\theta(z_\theta), z'_\xi \rangle }{||q_\theta(z_\theta)||_2 \cdot ||z'_\xi||_2},
\label{eq:eq-byol-mse}
\end{equation}
where $\langle\cdot, \cdot\rangle$ denotes the inner product.
To symmetrize the loss $L_{\theta,\xi}$, $L'_{\theta,\xi}$ is computed by feeding $v'$ to the online network and $v$ to the target network. The final loss is defined as $L_{\theta,\xi}^{\mathrm{BYOL}} = L_{\theta,\xi} + L'_{\theta,\xi}$.
At each training step, BYOL minimizes this loss function with respect to $\theta$ only, but $\xi$ is a slowly moving exponential average of $\theta$: $\xi \leftarrow \tau \xi + (1 - \tau) \theta$, where $\tau$ is a target decay rate.

It has been empirically shown that the combination of adding the predictor to the online network and using the moving average of the online network parameters as the target network encourages encoding more and more information within the online projection and avoids collapsed solutions such as constant representations.

\section{Proposed Method}

\begin{figure}[tb!]
  \centering
  \includegraphics[width=0.95\columnwidth]{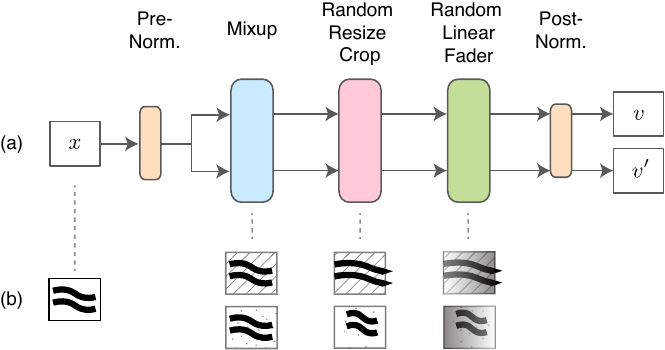} 
  \caption{Audio augmentation module of BYOL-A. (a) Block diagram. (b) Simplified example of outputs.}
  \label{fig:byola-aug}
\end{figure}

\begin{figure}[tb!]
  \centering
  \vspace{-30pt}
  \includegraphics[width=0.9\columnwidth]{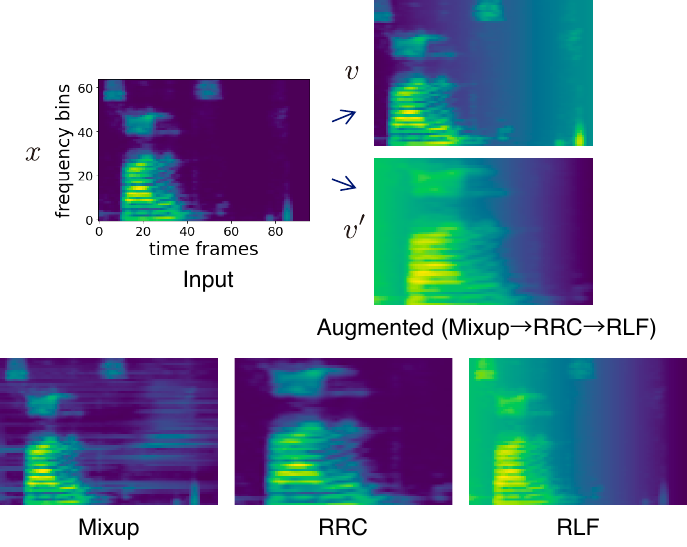} 
  \caption{Example of augmentation results. BYOL-A learns to cancel the difference between $v$ and $v'$. Each block result of the $v'$ is shown at the bottom.}
  \label{fig:byola-aug-examples}
  \vspace{-5pt}
\end{figure}

\subsection{BYOL for Audio (BYOL-A)}
We expand BYOL to learn representations of the input itself invariant to the perturbations of sounds by replacing data augmentations for audio, as shown in Fig. \ref{fig:byola-system}.

Fig. \ref{fig:byola-aug} shows the BYOL-A augmentation module composed of pre-/post-normalization blocks and three augmentation blocks: Mixup\cite{zhang2018mixup} (Section \ref{sec:byola-mixup}),
Random Resize Crop (RRC, Section \ref{sec:RRC}),
and Random Linear Fader (RLF, Section \ref{sec:RLF}).
Mixup makes random background sound, RRC makes random frequency/time shifts/stretches, and RLF makes random temporal amplitude changes, a simulation of random fade in/out.

Fig. \ref{fig:byola-aug-examples} shows an example of augmentations.
BYOL-A learns to cancel the difference between two augmentation results $v$ and $v'$.
As a result, the learned representations should be robust to the perturbation of sounds (e.g., robust to changes in background sound, pitch shift, time shift and stretch, and volume of sound).

The input raw audio samples are preprocessed into time-frequency (TF) features of the log-mel spectrogram, as has been done in previous studies. Then the module applies Pre-Norm, which normalizes input data $x$ to $\Tilde{x} = (x - \mu) / \sigma$, where $\mu$ and $\sigma$ are the average and standard deviation of training samples, respectively. This operation stabilizes the computations in the following augmentations blocks.
Similarly, after all the augmentations are applied, the module also applies Post-Norm, so that the final outputs of the augmentation module become $\sim {N}(0, 1)$. Augmentation operations can cause statistical drift in their outputs; the Post-norm corrects this possible drift.

As we describe in Section \ref{sec:byola-encoder}, we design the encoder for BYOL-A to encode representations to form multiple aspects of information by combining statistics of local and global features to meet various needs of tasks.

\subsubsection{Mixup for making background sound perturbation} \label{sec:byola-mixup}
We modify Mixup\cite{zhang2018mixup} or between-class (BC) learning\cite{tokozume2018between} to make slight randomness in background sound. These data augmentation techniques interpolate both features and labels between two data samples to create new data.
Using normalized log-mel spectrogram audio as input, our Mixup block randomly picks up a sample from a queue of past inputs and mixes it with the current input audio sample in a small ratio. As a result, mixed random audio becomes a part of the background sound in the current input.

While the original Mixup applies to both audio features and labels, our Mixup applies only to the features because we do not use labels.
In addition, as audio is log-scaled, we convert the input to a linear scale before the Mixup calculation and restore it to a log-scale again. In this paper, we refer to these operations as log-mixup-exp, from the analogy to the log-sum-exp\cite{logsumexp} calculation. Log-mixup-exp of $i$th input $x_i$ is

\vspace{-5pt}
\begin{equation}
\Tilde{x}_i = \log{\left((1 - \lambda) \exp(x_i) + \lambda \exp(x_k)\right)}
\label{eq:eq-log-mixup}
\end{equation}
where $x_k$ is a mixing counterpart, and mixing ratio $\lambda$ is sampled from the uniform distribution ${U}(0.0, \alpha)$, like in between-class learning.
In addition, $\alpha$ is a mixing ratio hyper-parameter that controls the degree of contrast between the resulting two mixed outputs. We observed that the evaluation result improves with smaller $\alpha$, $0.2$ for example, where $\Tilde{x}_i$ retains more of the original contents $x_i$ than its counterpart $x_k$ does, as we found in preliminary experiments.

$x_k$ is randomly chosen from a FIFO queue storing past inputs. As input is randomly sampled from the training dataset, samples in the queue form random subsets of the dataset. We store $2,048$ samples in the queue, which is larger than the batch size and big enough to maintain randomness.


\begin{figure}[tb!]
  \centering
  \includegraphics[width=0.8\columnwidth]{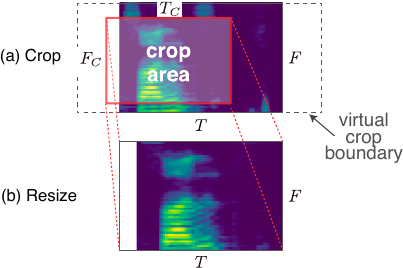} 
  \caption{Random resize crop of spectrogram. (a) Randomly chosen crop area content is (b) resized to the same size as the input.}
  \label{fig:byola-rrc}
\end{figure}

\subsubsection{Random Resize Crop (RRC) for making random frequency/time shift/stretch}\label{sec:RRC}
RRC is an image augmentation technique we use as an approximation of pitch shift and time shift and stretch of input log-mel spectrograms for learning representations invariant to the perturbations of frequency/time shift/stretch.

Fig.~\ref{fig:byola-rrc} shows the random crop procedure. The unit size of the input spectrogram consists of a number of frequency bins, $F$, and a number of time frames, $T$. First, we sample the random crop area from the virtual crop boundary, which has longer time frames than the input, $1.5 \times T$ for example. The size of the crop area is randomly sampled as
\begin{equation}
F_C = \lfloor\min({U}(f_1, f_2), 1.0) \times F\rfloor
\end{equation}
\begin{equation}
T_C = \lfloor {U}(t_1, t_2) \times T\rfloor
\end{equation}
where $F_C$ and $T_C$ are the number of frequency bins and number of time frames of random crop size, respectively, $f_1$ and $f_2$ form a frequency bin range $[f_1, f_2]$, $t_1$ and $t_2$ form a time frame range $[t_1, t_2]$, $\lfloor\cdot\rfloor$ is a floor function, and $\min(\cdot, \cdot)$ is a minimum function.
Contents in the crop area are then resized to the size of the input by bicubic interpolation. The virtual crop boundary is wider than the input, and we use $[0.6, 1.5]$ for both the frequency bin and time frame ranges in this paper, so the crop area can contain the outside of the input. This area is filled with zeros.
Note that we do not crop the outside of the frequency bin, which is restricted by the $\min()$ function in the $F_C$ calculation above.

\subsubsection{Random Linear Fader (RLF) for making random linear amplitude change}\label{sec:RLF}
RLF randomly applies a linear change in volume to the entire audio segment to simulate the approaching or passing away of a sound source, or fade-in/out. It adds random linear volume changes to the entire segment without losing the patterns of the contained sound events.

Let each element of the input spectrogram $x$ be $x[t,f]$, where $t$ is the time frame and $f$ is the frequency bin. First, we calculate temporal amplitude change $S[t]$ as follows:

\begin{equation}
S[t] = a + (b - a) \cdot t/T \quad \text{for } t \in \{0, ..., T - 1\},
\end{equation}
where $T$ is the number of time frames,
start frame gain $a \sim {U}(-1.0, 1.0)$, and end frame gain $b \sim {U}(-1.0, 1.0)$. The $S[t]$ is the gain for each time frame linearly interpolated from $a$ to $b$.
Then, we add $S$ to the input to make a linear amplitude change in log-mel spectrogram.

\begin{align}
  x'[t,f] = x[t,f] + S[t],\\
  \text{for } f \in \{0, ..., F - 1\},\notag\\
  \text{for } t \in \{0, ..., T - 1\}\notag
\end{align}
where $x'[t,f]$ is the result of RLF calculation, and $F$ is the number of frequency bins, respectively.
For example, if $a=-1$ and $b=0.5$, the relationship is $a<b$, which is an approximation of fade-in where the volume increases with time; if $a=0.5$ and $b=-0.5$, it is an approximation of fade-out where the volume decreases with time.

\subsubsection{BYOL-A encoder network}\label{sec:byola-encoder}
To enable a representation that provides multiple aspects of information, we make the BYOL-A encoder to (i) preserve all available information in global pooling, (ii) optimize the resolution of local features, (iii) combine local and global features, and (iv) combine average and maximum statistics in time.

We use the \textit{audio embedding block} from \cite{koizumi2020t6ntt} that satisfies requirement (i) as a base architecture. We make modifications to this base architecture to realize the remaining (ii)-(iv).

\begin{table}[tbph]
\caption{BYOL-A encoder architecture, an example of\\
input shape [B, 1, 64, 96] which is the size of [(B)atch, (C)hannel, (F)requency, (T)ime frame].}
\label{tab:byol-a-plus-encoder}
\centering
\begin{tabular}{lcl}
\toprule
Block & Output size & Operations \\
\midrule
Input & [B, 1, 64, 96] \\
\addlinespace[0.1cm]
Conv1 & [B, 64, 32, 48] & Conv2d 3x3 @ 64\\
& & BatchNorm2d @ 64 \\
& & ReLU \\
& & MaxPool2d 2x2, stride=2 \\
\addlinespace[0.1cm]
Conv2 & [B, 64, 16, 24] & Conv2d 3x3 @ 64\\
& & BatchNorm2d @ 64 \\
& & ReLU \\
& & MaxPool2d 2x2, stride=2 \\
\addlinespace[0.1cm]
Reshaping & [B, 24, 1024] & [B, C, F, T] to [B, T, D] \\
&& where $D=CF$\\
\addlinespace[0.1cm]
MLP & [B, 24, 2048] & FC 2048\\
 &  & ReLU\\
 &  & Dropout p=0.3\\
 &  & FC 2048\\
 &  & ReLU\\
\addlinespace[0.1cm]
Concat & [B, 24, 3072] & Concatenate outputs from\\
 &  & Reshaping and MLP\\
\addlinespace[0.1cm]
Pooling & [B, 3072] & \textit{Temporal mean+max pooling}\cite{kong2020panns}\\
\bottomrule
\end{tabular}
\end{table}

Table \ref{tab:byol-a-plus-encoder} shows the architecture, where 3x3 or 2x2 denotes the filter size, and the number after @ indicates the channel size, respectively.
This CNN takes the input to produce local features in Conv blocks, which is adjusted to make the receptive field smaller for (ii). Then, Reshaping flattens frequency and channel along the time axis, preserving available information to satisfy (i).
MLP learns to make useful global features on top of local features, and the following Concat concatenates both features to implement (iii). Finally, Pooling summarizes features into 3,072-d representation vectors using \textit{temporal mean+max pooling} to meet (iv).

To adapt the base CNN to our purposes, we made three modifications.
First, we reduce the number of convolutional blocks to increase the local feature resolution. One Conv block halves the output frequency and time resolution. We adjust the number of blocks to two, which reduces from three on the base architecture.
This adjustment directly changes the receptive field (RF) size, while tuning the RF is considered crucial for their generalization to unseen testing data\cite{koutini2021receptive}.
We conduct an ablation study in Section \ref{sec:exp-abl-byola-arch} and discuss more the necessity of RF adjustment in Appendix \ref{appendix:image-cnn-to-work}.

The second modification adds the Concat block that concatenates features from the earlier Reshaping block, a local feature, and later MLP block, a global feature.

The last modification adds the Pooling block that sums each element from temporal mean pooling and temporal max pooling of Concat output features, the \textit{temporal mean+max pooling}\cite{kong2020panns}, to accommodate advantages of both average and maximum statistics.

The encoder with these modifications, as a whole, make representations to combine local and global features as well as the statistics of the features while preserving frequency- and channel-wise information. The total number of encoder parameters is $6,333,376$.

\section{Experiments}\label{sec:experiments}

\begin{table*}[tbph]
\caption{Downstream task details.}
\label{tab:list-ds}
\centering
\begin{tabular}{rcccccccccc}\toprule
& \multicolumn{3}{c}{SER tasks} & \multicolumn{4}{c}{NOSS tasks} & \multicolumn{3}{c}{Music tasks} \\
\cmidrule(lr){2-4} \cmidrule(lr){5-8} \cmidrule(lr){9-11}  
 & ESC-50 &    US8K & FSD50K &    SPCV2 &    VC1 &     VF &    CRM-D &    GTZAN &     NSynth &      Surge \\
\midrule
\# of training samples & \multirow{3}{*}{\shortstack{5 folds\\ 2,000}} & \multirow{3}{*}{\shortstack{10 folds\\ 8,732}} & 36,796 & 84,843 & 138,361 & 121,281 & 5,155 & 443 & 289,205 & 148,896\\
\# of validation samples & & & 4,170 & 9,981 & 6,904 & 26,684 & 732 & 197 & 12,678 & 17,160 \\
\# of test samples & & & 10,231 & 11,005 & 8,251 & 28,463 & 1,551 & 290 & 4,096 & 17,336 \\
\# of classes & 50 & 10 & 200 & 35 & 1,251 & 6 & 6 & 10 & 11 & 88 \\
Average duration & 5.0 s & 4.0 s & 7.6 s & 1.0 s & 8.2 s & 5.8 s & 2.5 s & 30.0 s & 4.0 s &  4.0 s \\
\midrule
Previous studies
 & \cite{kong2020panns}\cite{gong2021psla}
 & \cite{guzhov2021esresnext} 
 & \cite{gong2021psla} 
 & \cite{gong2021ast}
 & \cite{shor2020trill} 
 & \cite{shor2020trill}
 & \cite{shor2020trill} 
 & \cite{kong2020panns} 
 & \cite{favory2020coala} 
 & (\cite{favory2020coala})$\ast$ \\
using tasks & \cite{guzhov2021esresnext}\cite{gong2021ast}\cite{cramer2019openl3}
 & \cite{favory2020coala}\cite{cramer2019openl3}
 & 
 & \cite{shor2020trill}\cite{saeed2020cola}
 & \cite{saeed2020cola}
 & \cite{saeed2020cola}
 & 
 & \cite{favory2020coala}
 & \cite{saeed2020cola}
 & (\cite{cramer2019openl3})$\ast$ \\
\bottomrule
\addlinespace[0.05cm]
\multicolumn{10}{l}{$^\ast$ Surge\cite{turian2021torchsynth} is a new dataset that evaluates itself using COALA\cite{favory2020coala} and OpenL3\cite{cramer2019openl3}.}\\
\end{tabular}
\end{table*}

To assess the generalizability of BYOL-A, we created a new benchmark described in Section \ref{sec:exp-benchmark} that covers a wide range of tasks, which we used throughout our experiments.
We detail the BYOL-A pre-training in Section \ref{sec:exp-byola-pretrain-details}.
Then, we evaluate BYOL-A and compare it with previous studies in Section \ref{sec:exp-eval-byola}.
We further conducted ablation studies: data augmentation block ablations in Section \ref{sec:exp-abl-byola}, network architecture ablations in Section \ref{sec:exp-abl-byola-arch}, global pooling ablations in Section \ref{sec:exp-abl-gp}, and BYOL framework ablations in Section \ref{sec:exp-abl-byol-frm}.
Lastly, we summarize the experiments in Section \ref{sec:exp-abl-sum}.

\subsection{Benchmark}\label{sec:exp-benchmark}
We explore general-purpose audio representations. 
To assess the generalizability of pre-trained models, we perform the standard linear evaluation protocol\cite{oord2018cpc, chen20simclr} using \textit{frozen} pre-trained models across a wide range of SER, NOSS, and music tasks collected from previous studies.

\subsubsection{Procedure details}

The linear evaluation pipeline first converts downstream task samples into feature embeddings using a \textit{frozen} pre-trained model as a feature extractor and then trains a linear layer with the task labels. It then gets the test results using the trained linear layer. All these results except FSD50K are accuracies; FSD50K results are mean average precision (mAP) and area under the curve (AUC).

All audio samples were randomly cropped to the average duration of the task dataset or added zero padding at the end, and they were resampled to the default sampling rate of each model.
For the models that come with a dedicated pre-processor that converts raw audio to TF feature, we used the pre-processor. The feature embeddings extracted by the models were standardized prior to training a linear layer.

We used the validation set for early stopping with a patience of 20 epochs and trained the linear layer for up to 200 epochs with the Adam optimizer. We manually tuned the learning rate to get the best results between 0.00001 and 0.01 for every test. We ran each evaluation three times and average the results.

\subsubsection{Downstream tasks}
We employed ten downstream tasks widely used in the previous studies as shown in Table \ref{tab:list-ds}: three sound event recognition (SER) tasks, four non-semantic speech\cite{shor2020trill} (NOSS) tasks, and three music tasks.
All tasks are multi-class single-label classifications except FSD50K, which is a multi-label classification. Therefore, we report FSD50K results separately. The following describes the tasks:

\begin{itemize}
    \item ESC-50\cite{piczak2015esc50}: a sound classification with 50 environmental sound classes. We conduct leave-one-out cross-validation (LOOCV) with the official five folds.
    \item UrbanSound8K\cite{salamon2014urbansound} (US8K): an urban sound classification task. We conduct LOOCV with the official ten folds.
    \item FSD50K\cite{fonseca2020fsd50k}: a multi-label classification SER task with 200 classes drawn from AudioSet ontology that covers a wide variety of sound events.
    \item Speech Commands V2\cite{speechcommandsv2} (SPCV2): a speech command word classification task, containing 105,829 utterances from 2,618 speakers.
    \item VoxCeleb1\cite{voxceleb} (VC1): a speaker identification task consisting of interviews of 1,251 celebrities.
    \item VoxForge\cite{voxforge} (VF): a language identification task: German, English, Spanish, French, Italian, and Russian.
    \item CREMA-D\cite{cao2014cremad} (CRM-D): a speech emotion recognition task of 91 speakers. We assign 70\% of speakers as training, 10\% as validation, and the remaining 20\% as test splits, with no speaker duplication in multiple splits.
    \item GTZAN\cite{gt2013gtzan}: a music genre recognition task. We follow fault-filtered partitioning\cite{kereliuk2015music}\cite{sturm2013gtzansplit}.
    \item NSynth\cite{nsynth2017}: an instrument family classification task of musical notes.
    \item Pitch Audio Dataset (Surge synthesizer)\cite{turian2021torchsynth} (Surge): a pitch audio classification task composed of 88 MIDI note classes and 2,084 tone preset sounds. We assign 10\% of the presets as validation, another 10\% as test, and the remaining 80\% as a training set.
\end{itemize}

\subsubsection{FSD50K extension: Sound event characteristic subsets}\label{sec:char-fsd50k}

In addition to the perspective of task diversity, we added to our benchmark a view of the sound event characteristics to gain an understanding of the utility of representations.
To do so, we introduced three subsets of FSD50K classes that group together the original classes that have similar characteristics of the sound events they contain. We report mAP results for each subset as well as the usual mAP results for all classes.
The following lists the subsets; Appendix \ref{appendix:CHAR-FSD50K} describes the details:

\begin{itemize}
    \item Single-source event: Class of a single sound source regardless of its repetition or continuation. 
    Examples:\\ 'Accordion', 'Bark', 'Bicycle bell', 'Boom', 'Clapping', 'Cough', 'Glass', 'Knock', 'Stream', 'Wind'.
    \item Sequential event: Class of a specific sequential sound occurrence of a sound source, characterized by the sequential change in pitch, amplitude, and/or timbre. 
    Examples: 'Car passing by', 'Fill (with liquid)', 'Hammer', 'Idling', 'Laughter', 'Ocean', 'Pour', 'Run', 'Tick-tock'.
    \item Scene event: Class of a group of multiple sound sources that describe a scene: 
    'Applause' 'Cheering', 'Crowd', 'Drum kit', 'Race car, auto racing', 'Subway, metro, underground'.
\end{itemize}

\subsection{Pre-training BYOL-A}\label{sec:exp-byola-pretrain-details}
We manually tuned hyperparameters for the BYOL framework and conducted an automatic parameter search for audio data augmentations to improve performance. The followings describe the details.

\subsubsection{BYOL framework settings}
We used the same MLPs in the original BYOL as the projection and prediction in BYOL-A networks, namely, a linear layer with an output size of $4,096$ followed by batch normalization (BatchNorm), rectified linear units (ReLU), and a linear layer to output embeddings with $256$ dimensions.
We trained for $100$ epochs with the Adam optimizer with a learning rate of $0.0001$, target decay rate parameter $\tau = 0.99$, and batch size of $256$.
While we tweaked the learning rate for better performance, we found that the default value of $\tau$ and the handy batch size pre-trains well. We further discuss this in Section \ref{sec:exp-abl-byol-frm}.

\subsubsection{Augmentation block parameters}
We conducted an exhaustive parameter search using Optuna\cite{akiba2019optuna} to achieve better performance in the pre-training.
As a result, we used the mixing ratio $\alpha$ of $0.2$ for Mixup; and for Random Resize Crop, we used the virtual crop boundary of $[F, 1.5 \times T]$ for the input size $[F, T]$, and both frequency bin/time frame ranges of $[0.6, 1.5]$.
We used a single set of augmentation $\mathcal{T}$, i.e., $t, t' \sim \mathcal{T}$.

\subsubsection{Dataset details}
We pre-trained using the $1,963,807$ samples (5,455 h) from balanced train and unbalanced train segments of the AudioSet\cite{gemmeke2017audioset} without labels.

In the ablation studies, we pre-trained using a development set of FSD50K\cite{fonseca2020fsd50k} without labels, $40,966$ samples (80 h) in total, with increased training epochs of $500$.

\subsection{Benchmarking BYOL-A and pre-trained models}\label{sec:exp-eval-byola}
To explore audio representations that generalize a wide range of tasks, we evaluate BYOL-A compared with various audio representations extracted from publicly available pre-trained models implementing previous methods.
These methods used different training frameworks, network architectures, and datasets, as described in Section \ref{sec:relworks-prev-models}; moreover, they shared benchmark tasks only partially, making comparison difficult.
With a unified benchmark, we can compare them and evaluate the generalizability of learned representations of the methods invented with diverse design choices.

\subsubsection{Representations from previous methods}
Table \ref{tab:list-ar} lists eleven audio representations from the eight pre-trained models we use.
We chose diverse state-of-the-art models that have evaluated different task performances.

\begin{table*}[tbph]
\caption{Details of audio representations extracted from pre-trained models.}
\label{tab:list-ar}
\centering
\begin{tabular}{lllllll}
\toprule
& \multicolumn{1}{c}{Pre-training} & \multicolumn{2}{c}{Encoder} & \multicolumn{2}{c}{Input} & \multicolumn{1}{c}{Output}\\
\cmidrule{2-2} \cmidrule(lr){3-4} \cmidrule(lr){5-6} \cmidrule(lr){7-7} 
Representation & dataset & architecture & prms. & feature & FS (Hz) & feature\\
\midrule
{[S]} VGGish\cite{hershey2017cnn} & YouTube-8M & VGGish (VGG\cite{vgg} based) & 72.1M & TF 64-d & 16,000 & 128-d/frame\\
{[S]} VGGish-4K\cite{hershey2017cnn} & YouTube-8M & VGGish (VGG\cite{vgg} based) & 72.1M & TF 64-d & 16,000 & 4,096-d/frame\\
{[Sas]} PANNs\cite{kong2020panns} & AudioSet & PANNs CNN14 & 79.7M & TF 64-d &    16,000  & 2,048-d\\
{[Sas]} ESResNeXt\cite{guzhov2021esresnext}  & ImageNet+AudioSet & ESResNe(X)t-fbsp (ResNeXt\cite{resnext} based) & 31.1M & TF(fbsp) 341-d &  44,100 & 2,048-d\\
{[Sas]} AST\cite{gong2021ast} & ImageNet+AudioSet & AST (ViT\cite{ViT}\cite{DeiT} based) & 87.0M & TF 128-d & 16,000 & 768-d\\
{[Ux]} COALA\cite{favory2020coala}  & Freesound & Audio encoder CNN & 2.4M & TF 96-d & 22,000 & 1,152-d/frame\\
{[Ux]} OpenL3-E\cite{cramer2019openl3}  &  AudioSet (environ.) & Audio sub network of $L^3$-Net\cite{Arandjelovic2017l3net} & 4.7M &  TF 256-d & 48,000 & 6,144-d/frame\\
{[Ux]} OpenL3-M\cite{cramer2019openl3} & AudioSet (music) & Audio sub network of $L^3$-Net\cite{Arandjelovic2017l3net} & 4.7M & TF 256-d & 48,000 & 6,144-d/frame\\
{[U]} TRILL\cite{shor2020trill} & AudioSet (speech) & ResNets/MobileNets & 9.0M/1.6M & TF 64-d & 16,000 & 12,288-d/frame\\
{[U]} Wav2Vec2-F\cite{baevski2020wav2vec2} & Librispeech & CNN & 4.2M & Raw audio & 16,000 & 512-d/frame\\
{[U]} Wav2Vec2-C\cite{baevski2020wav2vec2} &   Librispeech & CNN+Transformer & 315.4M & Raw audio & 16,000 & 1,024-d/frame\\
\midrule
{[U]} BYOL-A & AudioSet & CNN & 6.3M & TF 64-d & 16,000 & 3,072-d\\
\bottomrule
\end{tabular}
\end{table*}

We prefix the name of representations with labels: [Sas] for AudioSet-supervised learning, [S] for supervised learning on other datasets, [U] for audio unsupervised learning, and [Ux] for cross-modal unsupervised learning among audio and other modalities.

We extracted a single embedding per variable-length audio input. For some models that output embeddings frame by frame for input, we applied \textit{temporal mean+max pooling} except for Wav2Vec2-C so that we could make fair comparisons with BYOL-A. For Wav2Vec2-C, we temporally averaged their embeddings for making their best performance.

\begin{itemize}
    \item {[S]} VGGish and {[S]} VGGish-4K are from VGGish\cite{hershey2017cnn} pre-trained on YouTube-8M\cite{youtube8m}. {[S]} VGGish is the output 128-d per frame embeddings used by \cite{sethi2020soundscapes}\cite{koizumi2020tfmcaptioning}\cite{oncescu2021audioretrieval}\cite{kim2019audiocaps}\cite{tolkova2021parsing}, and {[S]} VGGish-4K is the 4,096-d per frame embeddings from the first FC block. 
    We use a PyTorch implementation of VGGish\footnote{\scriptsize \url{https://github.com/tcvrick/audioset-vggish-tensorflow-to-pytorch}}.
    \item {[Sas]} PANNs is the CNN14 from PANNs\footnote{\scriptsize \url{https://github.com/qiuqiangkong/audioset_tagging_cnn}}\cite{kong2020panns} pre-trained on AudioSet. We use 2,048-d embeddings that are the input to the last FC layer.
    \item {[Sas]} ESResNe(X)t-fbsp\footnote{\scriptsize \url{https://github.com/AndreyGuzhov/ESResNeXt-fbsp}}\cite{guzhov2021esresnext} (ESResNeXt) is a CNN pre-trained on ImageNet and AudioSet. We extract 2,048-d embeddings that are the input to the last FC layer.
    \item {[Sas]} AST\footnote{\scriptsize \url{https://github.com/YuanGongND/ast}}\cite{gong2021ast} is an Audio Spectrogram Transformer pre-trained on ImageNet and AudioSet. We use 768-d {[CLS]} token embeddings.
    \item {[Ux]} COALA\footnote{\scriptsize \url{https://github.com/xavierfav/coala}}\cite{favory2020coala} is a cross-modal self-supervised learning of audio and tags, pre-trained on a smaller dataset consisting of 170,793 training and 19,103 validation sounds from Freesound\cite{freesound}. We use the audio encoder output, 1,152-d per frame embeddings.
    \item {[Ux]} OpenL3-E and {[Ux]} OpenL3-M are from OpenL3\cite{cramer2019openl3}, cross-modal self-supervised learning of audio and video that pre-trains the $L^3$-Net\cite{Arandjelovic2017l3net} models on the environmental or music subset of AudioSet. We use 6,144-d per frame embeddings outputs from $L^3$-Nets audio encoders of a PyTorch implementation\footnote{\scriptsize \url{https://github.com/torchopenl3/torchopenl3}}.
    \item {[U]} TRILL\cite{shor2020trill} is a speech representation model pre-trained on speech containing clips from AudioSet. We use \textit{layer19} 12,288-d embeddings and pre-trained weights on the TensorFlow Hub\footnote{\scriptsize \url{https://tfhub.dev/google/nonsemantic-speech-benchmark/trill/2}}.
    \item {[U]} Wav2Vec2-F and {[U]} Wav2Vec2-C are from Wav2Vec2.0\cite{baevski2020wav2vec2}. {[U]} Wav2Vec2-F is the 512-d embeddings from the front-end CNN feature encoder, and {[U]} Wav2Vec2-C is the 1,024-d embeddings from the Transformer context network output. We use the Wav2Vec2-Large-960h-Lv60 pre-trained weights on the Hugging Face model hub\footnote{\url{https://huggingface.co/facebook/wav2vec2-large-960h-lv60}} pre-trained on Librispeech\cite{Panayotov2015LibrispeechAA}. Since it is pre-trained on the speech corpus only, downstream tasks other than speech context are challenging.
\end{itemize}

\subsubsection{Results and discussions}\label{sec:exp-9-ds-result-discussion}

\begin{table*}[htbp]
\caption{Linear evaluation benchmark accuracies (\%) with 95\% confidence intervals (CI) of audio representations on downstream tasks. BYOL-A is pre-trained on AudioSet. Top: reference results, middle \& bottom: results of this evaluation.}
\label{tab:result-benchmark}
\centering
\resizebox{\textwidth}{!}{%
\begin{tabular}{lllllllllll}\toprule
&  \multicolumn{2}{c}{SER tasks} & \multicolumn{4}{c}{NOSS tasks} & \multicolumn{3}{c}{Music tasks} & \\
\cmidrule(lr){2-3} \cmidrule(lr){4-7} \cmidrule(lr){8-10}  
Representation &    ESC-50 &    US8K &    SPCV2 &    VC1 &     VF &    CRM-D &    GTZAN &     NSynth &      Surge & Average \\
\midrule
{[U]} TRILL{\cite{shor2020trill}}    &   N/A &   N/A &  74.9 &  17.9 &  88.1 &  68.1$^\ast$ &   N/A &   N/A &   N/A &   N/A \\
{[U]} COLA{\cite{saeed2020cola}}   &   N/A &   N/A &  62.4 &  29.9 &  71.3 &   N/A &   N/A &  63.4 &   N/A &   N/A \\
{[Ux]} Wang et al.{\cite{wang2021multimodal}} &   N/A &   N/A &  82.2 &  38.2 &  79.0 &   N/A &   N/A &  68.3 &   N/A &   N/A \\
\midrule
{[S]} VGGish\cite{hershey2017cnn}      &  68.2 {\fontsize{6pt}{6pt}\selectfont $\pm$ 1.1} &  75.1 {\fontsize{6pt}{6pt}\selectfont $\pm$ 0.3} &  14.3 {\fontsize{6pt}{6pt}\selectfont $\pm$ 0.3} &   9.0 {\fontsize{6pt}{6pt}\selectfont $\pm$ 0.2} &  75.7 {\fontsize{6pt}{6pt}\selectfont $\pm$ 0.3} &  44.4 {\fontsize{6pt}{6pt}\selectfont $\pm$ 1.1} &  75.3 {\fontsize{6pt}{6pt}\selectfont $\pm$ 6.7} &  53.9 {\fontsize{6pt}{6pt}\selectfont $\pm$ 0.4} &   8.8 {\fontsize{6pt}{6pt}\selectfont $\pm$ 0.2} & 47.2\\
{[S]} VGGish-4K\cite{hershey2017cnn}   &  79.5 {\fontsize{6pt}{6pt}\selectfont $\pm$ 0.4} &  78.5 {\fontsize{6pt}{6pt}\selectfont $\pm$ 0.3} &  47.0 {\fontsize{6pt}{6pt}\selectfont $\pm$ 0.6} &  25.0 {\fontsize{6pt}{6pt}\selectfont $\pm$ 1.1} &  85.4 {\fontsize{6pt}{6pt}\selectfont $\pm$ 0.2} &  50.8 {\fontsize{6pt}{6pt}\selectfont $\pm$ 1.8} &  70.2 {\fontsize{6pt}{6pt}\selectfont $\pm$ 8.6} &  68.8 {\fontsize{6pt}{6pt}\selectfont $\pm$ 0.5} &  19.7 {\fontsize{6pt}{6pt}\selectfont $\pm$ 0.1} & 58.3\\
{[Sas]} PANNs\cite{kong2020panns}     &  90.1 {\fontsize{6pt}{6pt}\selectfont $\pm$ 0.4} &  82.0 {\fontsize{6pt}{6pt}\selectfont $\pm$ 0.7} &  51.4 {\fontsize{6pt}{6pt}\selectfont $\pm$ 0.3} &   8.0 {\fontsize{6pt}{6pt}\selectfont $\pm$ 0.4} &  75.0 {\fontsize{6pt}{6pt}\selectfont $\pm$ 0.4} &  50.7 {\fontsize{6pt}{6pt}\selectfont $\pm$ 0.2} &  79.7 {\fontsize{6pt}{6pt}\selectfont $\pm$ 3.1} &  66.0 {\fontsize{6pt}{6pt}\selectfont $\pm$ 0.3} &  10.4 {\fontsize{6pt}{6pt}\selectfont $\pm$ 0.7} & 57.0\\
{[Sas]} ESResNeXt\cite{guzhov2021esresnext} &  89.0 {\fontsize{6pt}{6pt}\selectfont $\pm$ 1.2} &  84.3 {\fontsize{6pt}{6pt}\selectfont $\pm$ 0.4} &  68.0 {\fontsize{6pt}{6pt}\selectfont $\pm$ 0.9} &  17.7 {\fontsize{6pt}{6pt}\selectfont $\pm$ 0.4} &  82.6 {\fontsize{6pt}{6pt}\selectfont $\pm$ 0.9} &  57.1 {\fontsize{6pt}{6pt}\selectfont $\pm$ 1.0} &  81.3 {\fontsize{6pt}{6pt}\selectfont $\pm$ 1.3} &  69.5 {\fontsize{6pt}{6pt}\selectfont $\pm$ 0.8} &  18.2 {\fontsize{6pt}{6pt}\selectfont $\pm$ 1.2} & 63.1\\
{[Sas]} AST\cite{gong2021ast}       &\textbf{93.5 {\fontsize{6pt}{6pt}\selectfont $\pm$ 0.4}}&\textbf{85.5 {\fontsize{6pt}{6pt}\selectfont $\pm$ 0.2}}&  71.8 {\fontsize{6pt}{6pt}\selectfont $\pm$ 0.4} &  16.5 {\fontsize{6pt}{6pt}\selectfont $\pm$ 0.4} &  81.2 {\fontsize{6pt}{6pt}\selectfont $\pm$ 0.2} &  57.9 {\fontsize{6pt}{6pt}\selectfont $\pm$ 0.6} &\textbf{84.3 {\fontsize{6pt}{6pt}\selectfont $\pm$ 1.8}}&  73.2 {\fontsize{6pt}{6pt}\selectfont $\pm$ 0.2} &  25.8 {\fontsize{6pt}{6pt}\selectfont $\pm$ 0.2} & 65.5\\
{[Ux]} COALA\cite{favory2020coala}      &  74.7 {\fontsize{6pt}{6pt}\selectfont $\pm$ 1.3} &  71.9 {\fontsize{6pt}{6pt}\selectfont $\pm$ 1.0} &  56.6 {\fontsize{6pt}{6pt}\selectfont $\pm$ 0.3} &  12.1 {\fontsize{6pt}{6pt}\selectfont $\pm$ 0.4} &  73.9 {\fontsize{6pt}{6pt}\selectfont $\pm$ 0.2} &  49.3 {\fontsize{6pt}{6pt}\selectfont $\pm$ 0.3} &  58.3 {\fontsize{6pt}{6pt}\selectfont $\pm$ 5.4} &  71.3 {\fontsize{6pt}{6pt}\selectfont $\pm$ 1.1} &  29.5 {\fontsize{6pt}{6pt}\selectfont $\pm$ 0.1} & 55.3\\
{[Ux]} OpenL3-E\cite{cramer2019openl3}   &  81.2 {\fontsize{6pt}{6pt}\selectfont $\pm$ 1.3} &  80.7 {\fontsize{6pt}{6pt}\selectfont $\pm$ 0.4} &  86.8 {\fontsize{6pt}{6pt}\selectfont $\pm$ 0.2} &  40.1 {\fontsize{6pt}{6pt}\selectfont $\pm$ 0.9} &  88.8 {\fontsize{6pt}{6pt}\selectfont $\pm$ 0.2} &  59.6 {\fontsize{6pt}{6pt}\selectfont $\pm$ 1.4} &  72.9 {\fontsize{6pt}{6pt}\selectfont $\pm$ 0.5} &  74.0 {\fontsize{6pt}{6pt}\selectfont $\pm$ 0.8} &\textbf{38.0 {\fontsize{6pt}{6pt}\selectfont $\pm$ 0.2}}& 69.1\\
{[Ux]} OpenL3-M\cite{cramer2019openl3}   &  82.2 {\fontsize{6pt}{6pt}\selectfont $\pm$ 0.8} &  80.4 {\fontsize{6pt}{6pt}\selectfont $\pm$ 0.3} &  87.9 {\fontsize{6pt}{6pt}\selectfont $\pm$ 0.1} &  40.7 {\fontsize{6pt}{6pt}\selectfont $\pm$ 0.6} &  90.1 {\fontsize{6pt}{6pt}\selectfont $\pm$ 0.6} &  60.4 {\fontsize{6pt}{6pt}\selectfont $\pm$ 1.0} &  73.3 {\fontsize{6pt}{6pt}\selectfont $\pm$ 3.0} &\textbf{75.6 {\fontsize{6pt}{6pt}\selectfont $\pm$ 0.5}}&  36.4 {\fontsize{6pt}{6pt}\selectfont $\pm$ 0.6} & 69.7\\
{[U]} TRILL\cite{shor2020trill}       &  75.4 {\fontsize{6pt}{6pt}\selectfont $\pm$ 0.7} &  75.2 {\fontsize{6pt}{6pt}\selectfont $\pm$ 1.3} &  78.4 {\fontsize{6pt}{6pt}\selectfont $\pm$ 0.8} &  40.1 {\fontsize{6pt}{6pt}\selectfont $\pm$ 1.1} &  88.8 {\fontsize{6pt}{6pt}\selectfont $\pm$ 0.3} &  58.8 {\fontsize{6pt}{6pt}\selectfont $\pm$ 2.3} &  64.4 {\fontsize{6pt}{6pt}\selectfont $\pm$ 1.8} &  74.3 {\fontsize{6pt}{6pt}\selectfont $\pm$ 1.8} &  28.7 {\fontsize{6pt}{6pt}\selectfont $\pm$ 1.0} & 64.9\\
{[U]} Wav2Vec2-F\cite{baevski2020wav2vec2}  &  65.6 {\fontsize{6pt}{6pt}\selectfont $\pm$ 1.7} &  67.8 {\fontsize{6pt}{6pt}\selectfont $\pm$ 0.3} &  85.8 {\fontsize{6pt}{6pt}\selectfont $\pm$ 0.2} &  32.0 {\fontsize{6pt}{6pt}\selectfont $\pm$ 0.3} &  81.7 {\fontsize{6pt}{6pt}\selectfont $\pm$ 0.1} &  56.4 {\fontsize{6pt}{6pt}\selectfont $\pm$ 0.5} &  62.3 {\fontsize{6pt}{6pt}\selectfont $\pm$ 1.0} &  62.2 {\fontsize{6pt}{6pt}\selectfont $\pm$ 0.8} &  30.0 {\fontsize{6pt}{6pt}\selectfont $\pm$ 0.4} & 60.4\\
{[U]} Wav2Vec2-C\cite{baevski2020wav2vec2}  &  57.6 {\fontsize{6pt}{6pt}\selectfont $\pm$ 0.8} &  66.9 {\fontsize{6pt}{6pt}\selectfont $\pm$ 0.4} &\textbf{96.6 {\fontsize{6pt}{6pt}\selectfont $\pm$ 0.0}}&  40.9 {\fontsize{6pt}{6pt}\selectfont $\pm$ 0.6} &\textbf{99.2 {\fontsize{6pt}{6pt}\selectfont $\pm$ 0.1}}&\textbf{65.5 {\fontsize{6pt}{6pt}\selectfont $\pm$ 1.7}}&  57.8 {\fontsize{6pt}{6pt}\selectfont $\pm$ 1.3} &  56.6 {\fontsize{6pt}{6pt}\selectfont $\pm$ 0.6} &  15.2 {\fontsize{6pt}{6pt}\selectfont $\pm$ 0.9} & 61.8\\
\midrule
{[U]} BYOL-A      &  83.2 {\fontsize{6pt}{6pt}\selectfont $\pm$ 0.6} &  79.7 {\fontsize{6pt}{6pt}\selectfont $\pm$ 0.5} &  93.1 {\fontsize{6pt}{6pt}\selectfont $\pm$ 0.4} &\textbf{57.6 {\fontsize{6pt}{6pt}\selectfont $\pm$ 0.2}}&  93.3 {\fontsize{6pt}{6pt}\selectfont $\pm$ 0.3} &  63.8 {\fontsize{6pt}{6pt}\selectfont $\pm$ 1.0} &  70.1 {\fontsize{6pt}{6pt}\selectfont $\pm$ 3.6} &  73.1 {\fontsize{6pt}{6pt}\selectfont $\pm$ 0.8} &  37.6 {\fontsize{6pt}{6pt}\selectfont $\pm$ 0.3} & \textbf{72.4}\\
\bottomrule
\addlinespace[0.05cm]
\multicolumn{11}{l}{$^\ast$ Reference result of intra-speaker task, trained and tested on one speaker at a time, then results averaged across speakers.}
\end{tabular}
}
\end{table*}

In Table \ref{tab:result-benchmark}, BYOL-A shows competitive results in all tasks with the best result of 72.4\% and the best VoxCeleb1 result of 57.6\%.

However, in the ESC-50, US8K, and GTZAN, BYOL-A has a performance gap compared to the AudioSet-supervised learning models. We think that AudioSet class supervision can cover similar class labels in these tasks.
For the SPCV2, VoxForge, and CREMA-D tasks, Wav2Vec2-C shows the best performance, suggesting that pre-training specialized for speech has advantages in spoken language tasks, while BYOL-A shows closer performance compared to other models.

Unsupervised learning models generally perform well in all tasks, suggesting that they effectively acquire general-purpose representations.
While TRILL and Wav2Vec2, pre-trained only on speech data, do not perform well in tasks other than speech, OpenL3-M, pre-trained on music samples, and OpenL3-E, pre-trained on environmental sounds, showed stable and good performance in tasks beyond the training data domain.

\begin{figure}[tb]
  \centering
  \includegraphics[width=0.9\columnwidth] {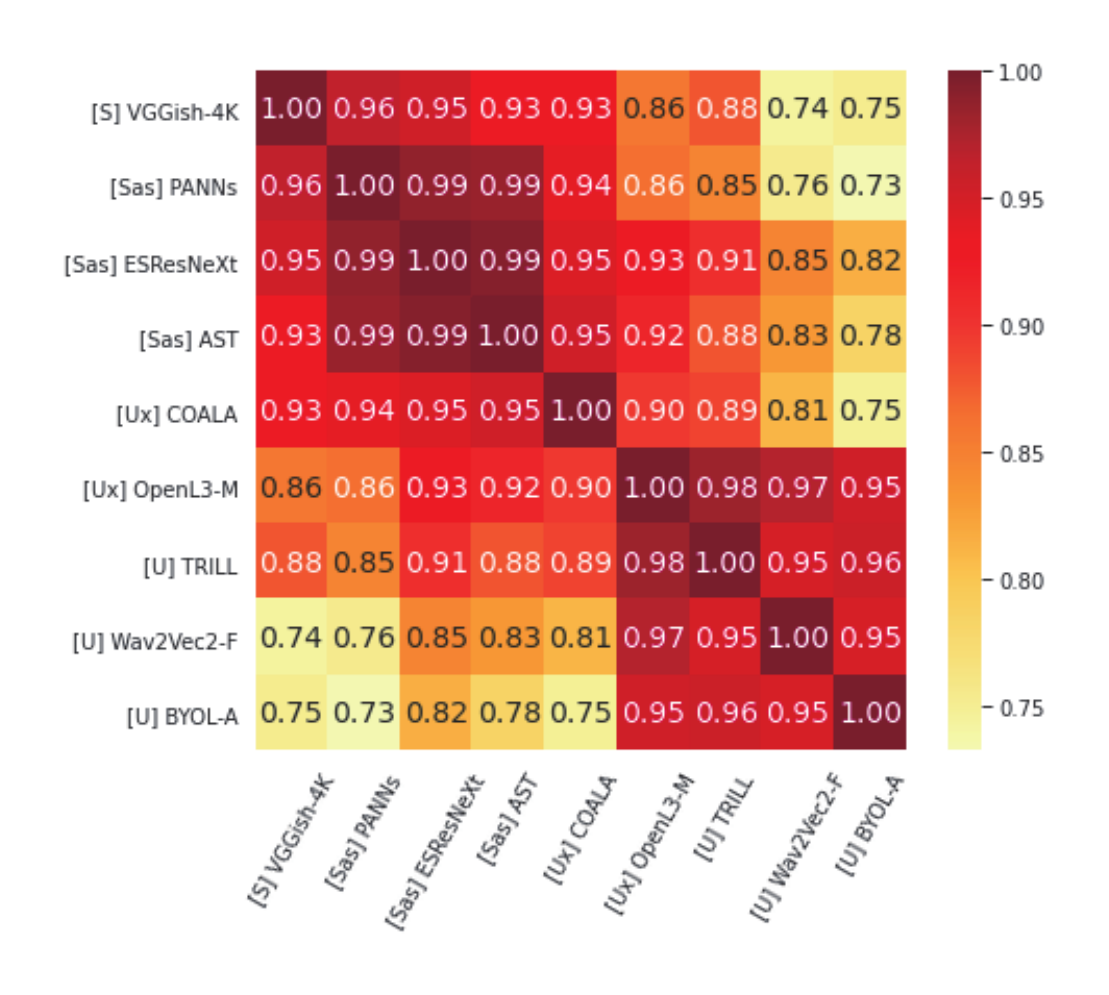}
  \caption{Pearson correlations coefficient among the result of representations in Table \ref{tab:result-benchmark}, excluding VGGish, OpenL3-E, and Wav2Vec2-C for brevity, providing performance trend similarity across representations. This highlights the different trends between supervised and unsupervised methods.}
  \label{fig:ar-corr-hm}
\end{figure}

Fig. \ref{fig:ar-corr-hm} shows the Pearson correlation coefficient across the result of representations in Table \ref{tab:result-benchmark}, suggesting several trends.
First, the correlations are high among the supervised learning methods and among the unsupervised learning methods, indicating that the label supervision tends to affect the task performance trend.
ESResNeXt and AST have particularly closer correlation trends, but while both are trained on ImageNet and AudioSet, they use different architectures (CNN and Transformer, respectively), suggesting that the performance trend can be more influenced by the learning method and dataset than the architecture.
COALA shows a performance trend similar to that of supervised learning, suggesting that cross modal training of tags and audio is similar to supervision of labels.

\subsubsection{Results on the FSD50K}\label{sec:exp-fsd50k}

Table \ref{tab:result-fsd50k} shows the results for FSD50K. Since the AudioSet class definition is a superset of FSD50K, we treat the [Sas] representations pre-trained on AudioSet as references.

\begin{table}[htbp]
\caption{FSD50K linear evaluation performances.}
\label{tab:result-fsd50k}
\centering
\begin{tabular}{llllll}
\toprule
{} &    \multicolumn{2}{c}{All classes} & \multicolumn{3}{c}{Char. subsets (mAP)} \\
\cmidrule(lr){2-3} \cmidrule(lr){4-6}
Representation &    (mAP) & (AUC) & Single & Seq.&  Scene \\
\midrule
{[Sas]} PANNs   $^\ast$  &  0.494 &  0.904 &  0.466 &      0.424 &  0.511 \\
{[Sas]} ESResNeXt $^\ast$&  0.524 &  0.911 &  0.490 &      0.443 &  0.611 \\
{[Sas]} AST     $^\ast$  &  0.589 &  0.928 &  0.561 &      0.528 &  0.609 \\
\midrule
{[S]} VGGish      &  0.320 &  0.886 &  0.287 &      0.260 &  \textbf{0.513} \\
{[S]} VGGish-4K   &  0.352 &  0.835 &  0.342 &      0.267 &  0.445 \\
{[Ux]} COALA      &  0.310 &  0.854 &  0.326 &      0.230 &  0.404 \\
{[Ux]} OpenL3-E   &  0.420 &  0.879 &  0.433 &      0.305 &  0.486 \\
{[Ux]} OpenL3-M   &  0.429 &  0.880 &  0.447 &      0.302 &  0.498 \\
{[U]} TRILL       &  0.330 &  0.841 &  0.336 &      0.241 &  0.455 \\
{[U]} Wav2Vec2-F  &  0.273 &  0.859 &  0.292 &      0.179 &  0.379 \\
{[U]} Wav2Vec2-C  &  0.240 &  0.843 &  0.252 &      0.200 &  0.332 \\
\midrule
{[U]} BYOL-A      &  \textbf{0.448} &  \textbf{0.896} &  \textbf{0.473} &      \textbf{0.313} &  0.496 \\
\bottomrule
\addlinespace[0.05cm]
\multicolumn{6}{l}{$^\ast$ Reference results for models pre-trained with AudioSet labels}\\
\multicolumn{6}{l}{\quad which is a super-set of FSD50K labels.}
\end{tabular}
\end{table}

FSD50K results show that OpenL3-E/M and BYOL-A achieve better results than the other representations, and BYOL-A performs the best, excluding the [Sas] representations.

The results in \textit{Char. subsets} columns, defined in Section \ref{sec:char-fsd50k}, show that these representations excel in detecting Single-source events, outperforming others with a large margin.
Whereas the performance gap between representations is smaller with Sequential and Scene events.
Considering that OpenL3-E/M and BYOL-A demonstrate better average results in Table \ref{tab:result-benchmark}, the results in Table \ref{tab:result-fsd50k} indicate that performing better with the Single-source event can lead to performing better in general audio tasks.

\subsection{Ablations of audio augmentation blocks}\label{sec:exp-abl-byola}
We assess the contribution of augmentation blocks by evaluating various combinations of these blocks in this section.

\begin{table}[tbhp]
\caption{Accuracies (\%) of BYOL-A augmentation block ablations.}
\vspace{-5pt}
\label{tab:result-augs}
\centering
\begin{tabular}{clll}\toprule
& Augmentation blocks used & Average & Diff.\\
\midrule
(a)& Mixup+RRC+RLF (BYOL-A) &\textbf{70.32}& \\
\midrule
(b)& Mixup+Gaussian+RRC+RLF &  70.12 &   -0.20 \\
(c)& Gaussian+RRC+RLF &  69.22 &   -1.10 \\
\addlinespace[0.1cm]
(d)& RRC+RLF &  69.46 &   -0.86\\
(e)& RRC+Mixup \cite{niizumi2021byol-a} &  70.30 &   -0.02 \\
\addlinespace[0.1cm]
(f)& RRC &   68.68 &   -1.64 \\
(g)& Mixup &  62.47 &   -7.85 \\
(h)& RLF &  57.70 &  -12.62 \\
\midrule
  & Random BYOL-A (no training) &  61.66 &  -8.66 \\
\bottomrule\\
\end{tabular}
\vspace{-8pt}
\end{table}

\begin{table*}[btp!]
\caption{Accuracies (\%) of BYOL-A encoder architectural ablations.}
\label{tab:result-abl-arch}
\centering
\vspace{-5pt}
\resizebox{\textwidth}{!}{%
\begin{tabular}{lllllllllllllll}\toprule
& & & [F, T] & \multicolumn{2}{c}{SER tasks} & \multicolumn{4}{c}{NOSS tasks} & \multicolumn{3}{c}{Music tasks} & &\\
\cmidrule(lr){5-8} \cmidrule(lr){7-10} \cmidrule(lr){11-13}  
Arch. & Dim. & Prms. & Reso. & ESC-50 &    US8K &    SPCV2 &    VC1 &     VF &    CRM-D &    GTZAN &     NSynth &      Surge & Avg. & Diff. \\
\midrule
(Base) BYOL-A Conv=2 & 3,072-d & 6.3M & [16, 24] &  82.5 &  78.8 &  \textbf{91.5} &  \textbf{51.4} &     91.4 &   58.5 &  65.1 &   75.5 &  \textbf{38.3} &    \textbf{70.3} &    \\
\midrule
BYOL-A Conv=1 & 3,072-d & 8.4M & [32, 48]  &  77.8 &  77.3 &  90.4 &  46.6 &     90.3 &   57.4 &  63.0 &   75.4 &  35.8 &    68.2 &  -2.1 \\
BYOL-A Conv=3 & 3,072-d & 5.3M  & [8, 12] &  83.2 &  78.4 &  90.3 &  43.2 &     90.7 &   61.6 &  65.5 &   74.6 &  31.1 &    68.7 &  -1.6 \\
\midrule
ResNet-18 (ReGP+N.RF)$^\ast$ & 4,096-d & 11.2M & [8, 6] &  \textbf{85.9} &  \textbf{79.7} &  90.3 &  38.5 &     90.7 &   63.5 &  69.5 &   75.5 &  27.2 &    69.0 &  -1.3 \\
ResNet-50 (ReGP+N.RF)$^\ast$ & 16,384-d & 23.5M & [8, 6] &  83.7 &  79.4 &  91.4 &  44.5 &     \textbf{93.1} &   \textbf{64.3} &  \textbf{71.8} &   \textbf{76.5} &  28.0 &    \textbf{70.3} &  0.0 \\
\bottomrule
\addlinespace[0.05cm]
\multicolumn{15}{l}{$^\ast$ Global pooling and receptive field (RF) size are modified. Appendix \ref{appendix:image-cnn-to-work} describes the details.}
\end{tabular}
}
\vspace{-5pt}
\end{table*}

\subsubsection{Experimental settings}
We tried combinations of Mixup, RRC, RLF, and an extra block, Gaussian, which interpolates training input with random data points sampled from the normal distribution. We added the Gaussian for comparison with Mixup.
The Gaussian block mixes the input with a random data point sampled from $\sim N(0, 0.4)$ using the log-mixup-exp calculation described in Section \ref{sec:byola-mixup}.

\subsubsection{Results and discussions}
Table \ref{tab:result-augs} shows the contribution of each augmentation and that combining them is essential for achieving the best performance of the BYOL-A. We also compare results with a randomly initialized model which is not pre-trained.

The single block results from (f) to (h) show that RRC is the most impactful, indicating that a representation ignoring the slight shifts/stretches in frequency/time axes is most effective for the downstream tasks.
In contrast, the (h) RLF result shows that pre-training only with a weak augmentation can impair the usefulness of a representation, even making the performance worse than a random model.
The results (d) RRC+RLF improve from the (f) RRC, showing that RLF can be useful if used with other blocks.
The final combination of (a) improved from (e) RRC+Mixup, our previous work\cite{niizumi2021byol-a}, by adding RLF.

Comparison between (a), (b), and (c) shows that Mixup, interpolating within-dataset samples, is more effective than Gaussian interpolating with random samples. 
The result of (a) Mixup+RRC+RLF is superior to that of (b) and (c), where (b) adds Gaussian on top of (a), and (c) replaces Mixup in (a) with Gaussian. In other words, mixing random noise cannot be as effective as mixing the sounds from the dataset for making background sound perturbations.

\subsection{Ablations of encoder network architecture}\label{sec:exp-abl-byola-arch}

We discuss architectural choices of the BYOL-A encoder for varying the number of convolutional blocks and even replacing the entire network with ResNet variants. Table \ref{tab:result-abl-arch} shows the results.

\subsubsection{Convolutional block ablations}

We compare three BYOL-A results in Table \ref{tab:result-abl-arch}, where we vary the number of convolutional blocks from one to three.

The BYOL-A Conv=1, a single Conv block with primitive output features with a rich resolution, results in the worst performance in most tasks. This result suggests that a single Conv block is insufficient to produce useful features.

The BYOL-A Conv=3, three Conv blocks with a lower resolution, degrade VC1 and Surge results. For solving VC1 (speaker identification) and Surge (pitch classification) tasks, frequency-wise information is considered important. Therefore, we think the degradation can be attributed to lower frequency resolution: $8$ (Conv=3) $<$ $16$ (BYOL-A, Conv=2).

\subsubsection{Replacing network with ResNets}
We compare the BYOL-A encoder CNN with two ResNet variants with modifications, which we describe the detail in Appendix \ref{appendix:image-cnn-to-work}. We used ResNet-18 and -50 as base ResNets.

Table \ref{tab:result-abl-arch} shows that BYOL-A is on par with the ResNet-50 variant and that it outperforms the ResNet-18 variant. 
Similar to BYOL-A Conv=3, ResNet variants show low results on VC1 and Surge while showing on par or better results on other tasks.
We think that the performance drop on VC1 and Surge is, as in BYOL-A Conv=3, due to lower frequency resolution; the variants have a resolution of 8, half of the BYOL-A's.
Making frequency strides smaller can increase the frequency resolution; however, it also increases the feature dimension. Doubling the ResNet-50 variant frequency resolution increases feature dimensions from 16,384-d to 32,768-d, making it closer to prohibitive for applications.

In summary, considering the trade-offs of the resolution, feature dimension, and model parameter size listed in the table, we choose BYOL-A with two convolutional blocks as a default encoder architecture that offers a balanced solution, which is also an improvement from our previous study\cite{niizumi2021byol-a}.
More modification on ResNet-50 or considering more sophisticated architectures such as Transformer variants could be good options. We leave them for future studies.

\subsection{Ablations of encoder global pooling blocks}\label{sec:exp-abl-gp}

We conducted an ablation study of the global pooling blocks in the BYOL-A encoder, namely the Reshaping, MLP, Concat, and Pooling blocks.
Tables \ref{tab:exp-abl-gp-confs} and \ref{tab:exp-abl-gp-results} show the configurations and corresponding results, respectively. We also conducted MLP size ablations, found in Table \ref{tab:result-abl-byol-mlp}.

\begin{table*}[htbp]
\centering
\caption{Configurations of BYOL-A encoder global pooling ablations.\\
Reshaping output is the size of (C)hannel and (F)requency along with (T)ime frame.}
\label{tab:exp-abl-gp-confs}
\vspace{-3pt}
\begin{tabular}{clcccccc}
\toprule
& & \multicolumn{4}{c}{Configurations}\\
\cmidrule(lr){3-6}
& & Reshaping & MLP & Concat & Temporal & Model & Output\\
& Remarks & output & block & block & pooling & prms. & dim.\\
\midrule
(1) & (Base) BYOL-A & $[T,CF]$ & Yes & Yes & mean+max & 6.3M & 3,072-d \\
\midrule
(2) & Frequency mean pooling $\leftarrow$ reshaping & $\mathbf{[T,C]}$ & Yes & Yes & mean+max & 4.4M & 2,112-d\\
(3) & Channel mean pooling $\leftarrow$ reshaping & $\mathbf{[T,F]}$ & Yes & Yes & mean+max & 4.3M & 2,064-d\\
\addlinespace[0.03cm]
\hdashline
\addlinespace[0.05cm]
(4) & Use global feature (MLP) only \cite{niizumi2021byol-a} & $[T,CF]$ & Yes & \textbf{No} & mean+max & 6.3M & 2,048-d \\
(5) & Use local feature (Reshaping) only &  $[T,CF]$ & \textbf{No} & Yes & mean+max & 0.38M & 1,024-d\\
\addlinespace[0.03cm]
\hdashline
\addlinespace[0.05cm]
(6) & Temporal mean pooling $\leftarrow$ maen+max & $[T,CF]$ & Yes & Yes & \textbf{mean} & 6.3M & 3,072-d \\
(7) & Temporal max pooling $\leftarrow$ maen+max & $[T,CF]$ & Yes & Yes & \textbf{max} & 6.3M & 3,072-d \\
\bottomrule
\end{tabular}
\vspace{-5pt}
\end{table*}

\begin{table*}[htbp]
\centering
\caption{Accuracies (\%) of BYOL-A encoder global pooling ablations, sub-categorized by conditions: (2,3) Freq./ch. pooling,\\
(4,5) Global/local features, and (6,7) Temporal pooling. We underline higher results in the sub category.}
\label{tab:exp-abl-gp-results}
\vspace{-3pt}
\begin{tabular}{cllllllllllll}\toprule
&  &  \multicolumn{2}{c}{SER tasks} & \multicolumn{4}{c}{NOSS tasks} & \multicolumn{3}{c}{Music tasks}\\
\cmidrule(lr){3-4} \cmidrule(lr){5-8} \cmidrule(lr){9-11}  
& Remarks &    ESC-50 &    US8K &    SPCV2 &    VC1 &     VF &    CRM-D &    GTZAN &     NSynth &      Surge & Average & Diff.\\
\midrule
(1)&(Base) BYOL-A &  82.5 &  78.8 &  91.5 &51.4&  91.4 &  58.5 &  65.1 & 75.5&  38.3 & 70.3\\
\midrule
(2)&Freq. mean pooling &  \underline{79.8} &  \underline{76.6} &  \underline{80.0} &  22.1 &  85.5 &  \underline{55.8} &  \underline{66.4} &  \underline{74.9} &  \underline{27.8} &  \underline{63.2} &  -7.1 \\
(3)&Channel mean pooling &  65.3 &  66.3 &  78.8 &  \underline{46.0} &  \underline{85.6} &  54.9 &  56.9 &  64.6 &  26.5 &  60.5 &  -9.8 \\
\addlinespace[0.03cm]
\hdashline
\addlinespace[0.05cm]
(4)&Global feature only \cite{niizumi2021byol-a} &\underline{83.6}&  \underline{79.0} &  \underline{90.6} &  \underline{44.4} &  \underline{91.1} &  \underline{57.9} &  \underline{64.7} &  \underline{73.5} &  29.8 &  \underline{68.3} &  -2.0 \\
(5)&Local feature only &  74.0 &  76.2 &  85.2 &  42.3 &  83.0 &  54.2 &  61.6 &  70.0 &  \underline{36.4} &  64.8 &  -5.6 \\
\addlinespace[0.03cm]
\hdashline
\addlinespace[0.05cm]
(6)&Temporal mean pooling &  73.7 &  75.5 &  88.5 &  45.3 &\underline{91.6}&\underline{58.5}&\underline{66.9}&  70.6 &\underline{38.9}&  67.7 &  -2.6 \\
(7)&Temporal max pooling  &  \underline{82.4} &\underline{79.5}&\underline{91.5}&  \underline{49.5} &  90.6 &  58.4 &  65.3 &  \underline{74.5} &  37.3 &  \underline{69.9} &  -0.4 \\
\bottomrule
\end{tabular}
\vspace{-5pt}
\end{table*}

Ablation results of Reshaping, (2) and (3) in Table \ref{tab:exp-abl-gp-results} indicate that averaging the frequency or channel deteriorates performance on downstream tasks.
We average the frequency or channel axis along time frames in these results.
The performance drop of (2) shows that averaging the frequency impairs results, especially on VC1 and Surge tasks where frequency-wise information is considered vital.
The result of (3) shows that averaging channels significantly degrade overall task performance, suggesting the importance of channel information to downstream tasks.
These averaging operations can be found in popular network architectures, even in audio models\cite{shor2020trill,kong2020panns,fonseca2020uclser20,guzhov2021esresnext}; however, these results indicate a potential negative performance impact in the previous studies.

The results (4) and (5) compare the performance difference between local and global features, namely, the MLP input and output features.
The results show that the global feature (4) increases accuracy on entire tasks except for Surge, showing that MLP learns useful features from the flattened frequency and channel information for each time frame while using most of the network capacity.
The result (1), concatenating both features, increases performance more than (4) global feature only in most tasks, indicating that many tasks benefit from both matured global and primitive local features.
Our previous version\cite{niizumi2021byol-a} was (4) global feature only, which we improved to (1) with the performance difference of 2.0.

Temporal pooling ablation results of (6) mean or (7) max pooling show that max pooling performs better than mean pooling on five tasks with a margin of 3.0 to 8.7\%, indicating that max pooling can be more advantageous in general. While the (1) combination of both statistics slightly degrades performance on some tasks, it improves on average, showing that tasks benefit from the combination of these statistics.

We also conducted an ablation study of MLP size, which is the output dimension of the FC layer in the MLP block. Table \ref{tab:result-abl-byol-mlp} shows that performance saturates at a size of 2,048, which we set as default in our encoder MLP.

\begin{table}[htbp]
\caption{Average accuracies (\%) of MLP size ablations.}
\label{tab:result-abl-byol-mlp}
\centering
\vspace{-5pt}
\begin{tabular}{llll}\toprule
MLP size & Model prms. &    Average & Difference \\ 
\midrule
8,192 & 75.6M          &    70.3 &   0.0 \\
4,096 & 21.0M          &    70.4 &   0.1 \\
2,048 (default) & 6.3M &    70.3 &  \\
1,024 & 2.1M           &    69.2 &  -1.1 \\
512 & 0.83M           &    67.7 &  -2.6 \\
\bottomrule
\end{tabular}
\vspace{-5pt}
\end{table}

\subsection{Ablations of BYOL framework}\label{sec:exp-abl-byol-frm}

To understand the contribution of the BYOL framework, we conducted ablation studies of its hyperparameters, namely, target decay rate $\tau$, which controls how close the target network weights become to the online network, and batch size, with which BYOL is reported to be robust.
We also additinally experimented removal of the prediction $q_\theta(z_\theta)$.

Target decay rate $\tau$ results in Table \ref{tab:result-abl-byol-prm-tau}  show that BYOL-A learns much more robustly with a target network than the original BYOL.
First, we can set the randomly-initialized BYOL-A result as a lower bound and the result of the pre-training with a default ($\tau = 0.99$) as a near upper bound; and we should see the results in between.
As the $\tau$ diverge from default $0.99$, such as $0.5$, $0.9$, and $0.999$, the performance degrades similarly as it does in the original BYOL. However, the degree of degradation is much smaller than in BYOL. We think this is because the lower bound of $61.7\%$ is very high compared to the original image BYOL of 1.4\%, reducing the room for degradation.

The $\tau=0.0$ and $1.0$ results indicate that BYOL-A can learn representations without a moving average target, even further, with a randomly initialized target.
The result of $\tau=1.0$, which fixes the random target weights, shows better results than the lower bound.
The result of $\tau=0.0$, which makes the target weight instantaneously update with the online's, shows a minor degradation of $-0.8$.
These results show that BYOL-A does not heavily rely on the the bootstrapping behavior of the BYOL framework to learn representations.

\begin{table}[htbp]
\vspace{-5pt}
\caption{Average accuracies (\%) of pre-training target decay rates $\tau$.}
\label{tab:result-abl-byol-prm-tau}
\centering
\vspace{-5pt}
\begin{tabular}{lll}\toprule
Target decay rate $\tau$ &    Average & Difference \\ 
\midrule
1.0 (fixed random init. target)   &    66.9 &  -3.4 \\
0.999 (slow moving target) &    69.7 &  -0.6 \\
0.99 (default) &    70.3 &  \\
0.9    &    70.0 &  -0.3 \\
0.5 (fast moving target) &   69.7 &  -0.6 \\
0.0 (target = online) &    69.5 &  -0.8 \\
\midrule
Random BYOL-A (no training) &    61.7 &  -8.6 \\
\bottomrule
\end{tabular}
\vspace{-3pt}
\end{table}

We further examined the necessity of the BYOL network components and confirmed that removing the prediction $q_\theta(z_\theta)$ broke BYOL-A learning, as shown in Table \ref{tab:result-abl-byol-nets}.
This result indicates that the standard network configuration of BYOL is essential for making BYOL-A viable, though it does not heavily depend on the bootstrapping of the target. 

\begin{table}[htbp]
\vspace{-5pt}
\caption{Average accuracies (\%) of framework component ablations}
\label{tab:result-abl-byol-nets}
\centering
\vspace{-5pt}
\begin{tabular}{llll}\toprule
\multicolumn{2}{l}{Configuration and corresponding loss} &  Average & Difference \\ 
\midrule
(Base) BYOL-A & $||\overline{q_\theta}(z_\theta) - \overline{z}'_\xi||^2_2$ &   70.3 &  \\
Target = online & $||\overline{q_\theta}(z_\theta) - \overline{z}'_\theta||^2_2$ &   69.5 &  -0.8 \\
No prediction & $||\overline{z}_\theta - \overline{z}'_\theta||^2_2$ &   52.8 &  -17.5 \\
\bottomrule
\end{tabular}
\end{table}

Batch size ablation results in Table \ref{tab:result-abl-byol-prm-bs} show that the performance does not degrade even with a small batch size of $64$.
The results other than the default of $256$ degraded slightly, but we think this can be attributed to the mismatch of the learning rate; we tested with a fixed learning rate that we optimized to the default batch size.
The results also empirically show that BYOL-A is more robust to the small batch size than the original BYOL, which showed degradation with a batch size of less than $256$, and more robust than COLA, a contrastive learning method that showed degradation with a batch size of less than $1024$.
We think this could also indicate that BYOL-A is less dependent on the inductive bias of the BYOL framework. 

\begin{table}[htbp]
\vspace{-5pt}
\caption{Average accuracies (\%) of pre-training batch size ablations.}
\label{tab:result-abl-byol-prm-bs}
\centering
\vspace{-5pt}
\begin{tabular}{lll}\toprule
Batch size &    Average & Difference \\ 
\midrule
1024   &    69.5 &  -0.8 \\
512    &    69.4 &  -0.9 \\
256 (default) &  70.3 &  \\
128    &    69.7 &  -0.6 \\
64     &    69.7 &  -0.6 \\
\bottomrule
\end{tabular}
\vspace{-5pt}
\end{table}

\subsection{Summary of experiments} \label{sec:exp-abl-sum}
We demonstrated the generalizability of the BYOL-A compared across major pre-trained models on a benchmark consisting of various tasks in Section \ref{sec:exp-eval-byola}.
In addition, intensive ablation studies provided evidence of contributions from different aspects.
To gain a holistic understanding of what makes the BYOL-A representation learning happen, we summarize the contribution of the components in Table \ref{tab:contributions}.

\begin{table}[htbp]
\vspace{-5pt}
\caption{Summary of performance contributions (\%).}
\label{tab:contributions}
\centering
\vspace{-5pt}
\begin{tabular}{lll}\toprule
Component/Ablation & Contribution & Reference\\ 
\midrule
Encoder architecture & +61.7 \\
- Feature reshaping along time & -9.8 to -7.1 & Table \ref{tab:exp-abl-gp-results} (2), (3) \\
- MLP & -5.6 & Table \ref{tab:exp-abl-gp-results} (5) \\
- Combining local\&global features & -2.0 & Table \ref{tab:exp-abl-gp-results} (4) \\
- Temporal mean+max pooling & -2.6 to -0.4 & Table \ref{tab:exp-abl-gp-results} (6) (7) \\
\midrule
BYOL framework + BYOL-A augs. & +8.6  \\
- BYOL: bootstrapping & -0.8 & Table \ref{tab:result-abl-byol-nets} \\
- BYOL: prediction $q_\theta(z_\theta)$ & -17.5 & Table \ref{tab:result-abl-byol-nets} \\
- Aug: Mixup\&RLF (use RRC only)  & -1.6 & Table \ref{tab:result-augs} (f) \\
- Aug: RRC\&Mixup (use RLF only)  & -12.6 & Table \ref{tab:result-augs} (h) \\
\bottomrule
\end{tabular}
\end{table}

Ablation results clarified that the BYOL-A encoder architecture contributes the most to performance, achieving 61.7\% with only random initialization.
Pre-training improves performance to 70.3\%, showing that the BYOL framework and BYOL-A audio data augmentations contribute the last +8.6.

We think the convolutional blocks are the most significant performance factor for the encoder, followed by feature reshaping along time, MLP, combining local \& global features, and temporal mean+max pooling.
We estimate the performance contribution of the convolutional blocks can be up to about +46, which is the average accuracy of the whole encoder minus the other factors, i.e., $61.7 - 7.1 - 5.6 - 2.0 - 0.4 \approx 46$.

The ablation results related to framework and augmentations show that both the BYOL framework and augmentations are crucial to making the performance improvement viable.
A simpler framework as such removing prediction $q_\theta(z_\theta)$ fails the training.
Similarly, using a noticeably weak augmentation such as RLF only also fails the training.
The BYOL framework and the BYOL-A audio data augmentations work together to achieve the performance of BYOL-A.

In summary, the inductive bias of the BYOL-A encoder network architecture primarily contributes to the performance of BYOL-A, and the pre-training under the BYOL framework with BYOL-A audio data augmentations completes the final critical portion of the performance.
As a whole, BYOL-A achieves the best average result among various pre-trained models, demonstrating its generalizability as a general-purpose audio representation.

\section{Conclusion}
In this study, we explored pre-trained audio representations for general audio tasks.
We hypothesized that representations effective for general audio tasks should provide multiple aspects of robust features of the input sound.
Robust features can help sound applications (i.e., recognition) under perturbations such as varying pitch or timbre. Representations providing multiple aspects of information calculated using these features can help various purposes.
As a result, these representations should serve to meet the diverse needs of tasks.

We proposed a self-supervised learning method called Bootstrap Your Own Latent (BYOL) for Audio (BYOL-A, pronounced "viola") to pre-train audio representations invariant to the slight perturbations of background sound, frequency/time shift/stretch, and temporal amplitude change. To make representations that provide multiple aspects of features, we made the BYOL-A encoder combine statistics of local and global features while preserving frequency- and channel-wise information.

We evaluated the general-purpose task performance among various previous state-of-the-art methods on a benchmark composed of ten SER, NOSS, and music tasks.
The BYOL-A demonstrated the generalizability of its representation with the best average performance of 72.4\% and the best VoxCeleb1 performance of 57.6\%.

Extensive ablation studies clarified the contributions of BYOL-A components. We found that a large portion of the performance comes from the inductive bias of the BYOL-A encoder network architecture, and that the final critical portion resorts to the BYOL framework and BYOL-A audio data augmentations.
As a whole, BYOL-A learns to produce effective representations that generalize to various tasks.

We make our code available online and hope it fosters progress in future studies of audio representations.




\appendices

\section{Making an image-CNN-based model perform on general audio task benchmark}\label{appendix:image-cnn-to-work}
To further elaborate on our encoder design described in Section \ref{sec:byola-encoder}, we discuss what makes a CNN model perform well on various tasks in our benchmark using an image-CNN-based architecture as an example.
We use a ResNet-18\cite{resnet} from image-CNNs, and we change its input channels from three to one and remove the FC layer. Then the modified version, named ResNet-like, accepts batch input with a shape [(B)atch, 1, (F)requency, (T)ime frame], and outputs [B, 512], 512-d embeddings.

We made two improved versions based on the ResNet-like.
One is 'ResNet-like (ReGP)', where a ResNet-like replaces global pooling (Replacement of Global Pooling; ReGP).
The other is 'ResNet-like (ReGP + Narrow RF)', which is a ResNet-like (ReGP) with a modification so that the receptive field (RF) becomes narrower.

ResNet-like has a global average pooling that averages frequency and time axes and outputs 512-d embeddings. ResNet-like (ReGP) replaces the global average pooling with the Reshaping and Pooling blocks from the BYOL-A encoder, making output as 1,024-d embeddings.

ResNet-like (ReGP + Narrow RF) adjusts RF by changing strides [2, 2, 2, 2, 2] to [1, 2, 2, 2, [1, 2]]. The [1, 2] changes the frequency stride only to 1. This modification will make the RF half on the time axis and on fourth on the frequency axis. With an input shape [64, 96], which is [Frequency, Time frame], the output shape is [2, 3], which becomes [8, 6] after the modification. This especially changes the frequency resolution from two to eight, accommodating more frequency information available on downstream tasks.
This increases the embedding size: $512 \text{ channel} \times 8 \text{ frequency bins} = 4,096$-d.

We pre-trained these models in BYOL-A by replacing the encoder with them, using the same setting as in the BYOL-A ablation studies described in Section \ref{sec:exp-byola-pretrain-details}.

Table \ref{tab:result-resnet-rf} shows the results of the ResNet-like variants.
The base model's performance, ResNet-like, is 61.4\%, and it improves with ResNet-like (ReGP) to 63.3\%.
The performance of ResNet-like (ReGP + Narrow RF) improves to 69.0\%, which is comparable to BYOL-A's, 70.3\%.

These results show that global pooling in the image-CNN-based architecture needs to be improved, as discussed in Section \ref{sec:global-pooling-designs}. Moreover, adjusting frequency resolution is also crucial for good performance in general audio tasks, as reported in the previous study \cite{koutini2021receptive}.

\begin{table}[htbp]
\vspace{-5pt}
\caption{Linear evaluation benchmark accuracies (\%) of ResNet-based models.}
\label{tab:result-resnet-rf}
\centering
\vspace{-5pt}
\begin{tabular}{ll}\toprule
ResNet variation &    Average \\
\midrule
ResNet-like (ReGP + Narrow RF) &\textbf{69.0}\\
ResNet-like (ReGP)   &  63.3 \\
ResNet-like      &  61.4 \\
\bottomrule
\end{tabular}
\vspace{-10pt}
\end{table}

\section{FSD50K sound event characteristic subsets detail}\label{appendix:CHAR-FSD50K}
This appendix describes assigning FSD50K classes to the three sound event characteristic subsets defined in Section \ref{sec:char-fsd50k}.
We conducted the following steps to examine all FSD50K classes and determined the assignment.
First, we randomly select 50 samples from the target class. Then, we conducted a manual inspection by listening to each sample to determine which subset the sample belongs to.
After inspecting all 50 samples, only if 80\% (40 samples) or more fall into one subset, we assigned the target class to the subset.
We excluded classes that fell into multiple subsets (e.g., the Liquid class falls into a single-source and sequential event) and classes with vague characteristics (e.g., Mechanisms, Wood, etc.) from the assignment.
We repeated these steps and finally assigned 93 classes to one of the subsets out of 200 classes.

Table \ref{tab:CHAR-FSD50K label list} lists the FSD50K classes assigned to the sound event characteristic subsets.

\begin{table*}[htbp]
\vspace{-5pt}
\caption{FSD50K sound event characteristic subsets and the list of assigned classes.}
\label{tab:CHAR-FSD50K label list}
\vspace{-5pt}
\centering
\begin{tabular}{lcl}
\hline
Subset & Count & FSD50K classes assigned to the subset\\
\hline
Single & 66 & 'Accordion' 'Acoustic guitar' 'Bark' 'Bass drum' 'Bass guitar'
 'Bicycle bell' 'Boom' 'Burping, eructation' 'Chink, clink'\\
source && 'Chuckle, chortle' 'Clapping' 'Coin (dropping)' 'Cough' 'Cowbell'
 'Crash cymbal' 'Crow' 'Cutlery, silverware' 'Cymbal' \\
&& 'Dishes, pots, and pans' 'Fart' 'Finger snapping' 'Gasp' 'Glass'
 'Glockenspiel' 'Gunshot, gunfire' 'Gurgling' 'Harp' \\
&& 'Hi-hat' 'Keyboard (musical)' 'Knock' 'Mallet percussion' 'Marimba, xylophone'
 'Meow' 'Percussion' 'Piano' \\
&& 'Plucked string instrument' 'Power tool' 'Rain' 'Raindrop' 'Rattle (instrument)' 'Ringtone' 'Scissors' \\
&& 'Scratching (performance technique)' 'Screaming' 'Shatter' 'Shout' 'Sigh' 'Slam' 'Snare drum' 'Speech' \\
&& 'Speech synthesizer' 'Stream' 'Tambourine' 'Tap' 'Tearing' 'Telephone' 'Thump, thud' 'Tools' 'Truck' 'Trumpet' \\
&& 'Water' 'Whoosh, swoosh, swish' 'Wild animals' 'Wind' 'Wind chime'
 'Zipper (clothing)' \\
Sequential & 21 & 'Accelerating, revving, vroom' 'Boiling' 'Car passing by' 'Crying, sobbing' 'Fill (with liquid)' 'Frying (food)' 'Giggle' \\
&& 'Hammer'
 'Idling' 'Laughter' 'Mechanical fan' 'Motorcycle' 'Ocean' 'Pour' 'Run' \\
&& 'Sawing' 'Sliding door' 'Tick-tock' 'Traffic noise, roadway noise'
 'Walk, footsteps' 'Yell' \\
Scene & 6 &  'Applause' 'Cheering' 'Crowd' 'Drum kit' 'Race car, auto racing' 'Subway, metro, underground' \\
\hline
\end{tabular}
\vspace{-10pt}
\end{table*}

\ifCLASSOPTIONcaptionsoff
  \newpage
\fi

\bibliographystyle{IEEEtran}
\bibliography{refs}

\begin{thebibliography}{10}
\providecommand{\url}[1]{#1}
\csname url@samestyle\endcsname
\providecommand{\newblock}{\relax}
\providecommand{\bibinfo}[2]{#2}
\providecommand{\BIBentrySTDinterwordspacing}{\spaceskip=0pt\relax}
\providecommand{\BIBentryALTinterwordstretchfactor}{4}
\providecommand{\BIBentryALTinterwordspacing}{\spaceskip=\fontdimen2\font plus
\BIBentryALTinterwordstretchfactor\fontdimen3\font minus
  \fontdimen4\font\relax}
\providecommand{\BIBforeignlanguage}[2]{{%
\expandafter\ifx\csname l@#1\endcsname\relax
\typeout{** WARNING: IEEEtran.bst: No hyphenation pattern has been}%
\typeout{** loaded for the language `#1'. Using the pattern for}%
\typeout{** the default language instead.}%
\else
\language=\csname l@#1\endcsname
\fi
#2}}
\providecommand{\BIBdecl}{\relax}
\BIBdecl

\bibitem{bert}
J.~Devlin, M.~Chang, K.~Lee, and K.~Toutanova, ``{BERT:} pre-training of deep
  bidirectional transformers for language understanding,'' in \emph{NAACL-HLT},
  2019, pp. 4171--4186.

\bibitem{vgg}
K.~Simonyan and A.~Zisserman, ``Very deep convolutional networks for
  large-scale image recognition,'' in \emph{ICLR}, 2015.

\bibitem{resnet}
K.~He, X.~Zhang, S.~Ren, and J.~Sun, ``Deep residual learning for image
  recognition,'' in \emph{CVPR}, 2016, pp. 770--778.

\bibitem{resnext}
S.~Xie, R.~Girshick, P.~Dollar, Z.~Tu, and K.~He, ``Aggregated residual
  transformations for deep neural networks,'' in \emph{CVPR}, Jul 2017.

\bibitem{hershey2017cnn}
S.~Hershey, S.~Chaudhuri, D.~P.~W. Ellis, J.~F. Gemmeke, A.~Jansen, R.~C.
  Moore, M.~Plakal, D.~Platt, R.~A. Saurous, B.~Seybold, M.~Slaney, R.~Weiss,
  and K.~Wilson, ``Cnn architectures for largescale audio classification,'' in
  \emph{ICASSP}, 2017, pp. 131--135.

\bibitem{koike2020heartsound}
T.~Koike, K.~Qian, Q.~Kong, M.~D. Plumbley, B.~W. Schuller, and Y.~Yamamoto,
  ``Audio for audio is better? an investigation on transfer learning models for
  heart sound classification,'' in \emph{EMBC}, 2020, pp. 74--77.

\bibitem{balagopalan2021alzheimer}
A.~Balagopalan and J.~Novikova, ``Comparing acoustic-based approaches for
  alzheimer's disease detection,'' \emph{arXiv preprint arXiv:2106.01555},
  2021.

\bibitem{sethi2020soundscapes}
S.~S. Sethi, N.~S. Jones, B.~D. Fulcher, L.~Picinali, D.~J. Clink, H.~Klinck,
  C.~D.~L. Orme, P.~H. Wrege, and R.~M. Ewers, ``Characterizing soundscapes
  across diverse ecosystems using a universal acoustic feature set,'' vol. 117,
  no.~29.\hskip 1em plus 0.5em minus 0.4em\relax National Academy of Sciences,
  2020, pp. 17\,049--17\,055.

\bibitem{koizumi2020tfmcaptioning}
Y.~Koizumi, R.~Masumura, K.~Nishida, M.~Yasuda, and S.~Saito, ``A
  transformer-based audio captioning model with keyword estimation,'' in
  \emph{Interspeech}, Oct 2020.

\bibitem{oncescu2021audioretrieval}
A.-M. Oncescu, A.~S. Koepke, J.~F. Henriques, Z.~Akata, and S.~Albanie, ``Audio
  retrieval with natural language queries,'' \emph{arXiv preprint
  arXiv:2105.02192}, 2021.

\bibitem{grill2020byol}
J.-B. Grill, F.~Strub, F.~Altch^^c3^^a9, C.~Tallec, P.~H. Richemond,
  E.~Buchatskaya, C.~Doersch, B.~A. Pires, Z.~D. Guo, M.~G. Azar, B.~Piot,
  K.~Kavukcuoglu, R.~Munos, and M.~Valko, ``Bootstrap your own latent - a new
  approach to self-supervised learning,'' in \emph{NeurIPS}, 2020.

\bibitem{kong2020panns}
Q.~Kong, Y.~Cao, T.~Iqbal, Y.~Wang, W.~Wang, and M.~D. Plumbley, ``Panns:
  Large-scale pretrained audio neural networks for audio pattern recognition,''
  \emph{IEEE/ACM Trans. Audio, Speech, Language Process.}, vol.~28, pp.
  2880--2894, 2020.

\bibitem{gong2021psla}
Y.~Gong, Y.-A. Chung, and J.~Glass, ``Psla: Improving audio tagging with
  pretraining, sampling, labeling, and aggregation,'' \emph{arXiv preprint
  arXiv:2102.01243}, 2021.

\bibitem{guzhov2021esresnext}
A.~Guzhov, F.~Raue, J.~Hees, and A.~Dengel, ``Esresne(x)t-fbsp: Learning robust
  time-frequency transformation of audio,'' in \emph{IJCNN}, Jul 2021.

\bibitem{gong2021ast}
Y.~Gong, Y.-A. Chung, and J.~Glass, ``Ast: Audio spectrogram transformer,''
  \emph{arXiv preprint arXiv:2104.01778}, 2021.

\bibitem{favory2020coala}
X.~Favory, K.~Drossos, T.~Virtanen, and X.~Serra, ``Coala: Co-aligned
  autoencoders for learning semantically enriched audio representations,'' in
  \emph{ICML}, Jul 2020.

\bibitem{cramer2019openl3}
J.~Cramer, H.-H. Wu, J.~Salamon, and J.~P. Bello, ``Look, listen and learn
  more: Design choices for deep audio embeddings,'' in \emph{ICASSP}, Brighton,
  UK, May 2019, pp. 3852--^^e2^^80^^933\,856.

\bibitem{wang2021multimodal}
L.~Wang, P.~Luc, A.~Recasens, J.-B. Alayrac, and A.~van~den Oord, ``Multimodal
  self-supervised learning of general audio representations,'' \emph{arXiv
  preprint arXiv:2104.12807}, 2021.

\bibitem{shor2020trill}
J.~Shor, A.~Jansen, R.~Maor, O.~Lang, O.~Tuval, F.~d.~C. Quitry,
  M.~Tagliasacchi, I.~Shavitt, D.~Emanuel, and Y.~Haviv, ``Towards learning a
  universal non-semantic representation of speech,'' in \emph{Interspeech}, Oct
  2020.

\bibitem{saeed2020cola}
A.~Saeed, D.~Grangier, and N.~Zeghidour, ``Contrastive learning of
  general-purpose audio representations,'' in \emph{ICASSP}, Jun 2021.

\bibitem{fonseca2020uclser20}
E.~Fonseca, D.~Ortego, K.~McGuinness, N.~E. O’Connor, and X.~Serra,
  ``Unsupervised contrastive learning of sound event representations,'' in
  \emph{ICASSP}, Jun 2021.

\bibitem{gemmeke2017audioset}
J.~F. Gemmeke, D.~P.~W. Ellis, D.~Freedman, A.~Jansen, W.~Lawrence, R.~C.
  Moore, M.~Plakal, and M.~Ritter, ``Audio set: An ontology and human-labeled
  dataset for audio events,'' in \emph{ICASSP}, 2017, pp. 776--780.

\bibitem{piczak2015esc50}
K.~J. Piczak, ``{ESC}: {Dataset} for {Environmental Sound Classification},'' in
  \emph{ACM-MM}, 2015, pp. 1015--1018.

\bibitem{salamon2014urbansound}
J.~Salamon, C.~Jacoby, and J.~P. Bello, ``A dataset and taxonomy for urban
  sound research,'' in \emph{ACM-MM}, Nov. 2014, pp. 1041--1044.

\bibitem{fonseca2020fsd50k}
E.~Fonseca, X.~Favory, J.~Pons, F.~Font, and X.~Serra, ``Fsd50k: an open
  dataset of human-labeled sound events,'' \emph{arXiv preprint
  arXiv:2010.00475}, 2020.

\bibitem{speechcommandsv2}
P.~{Warden}, ``{Speech Commands: A Dataset for Limited-Vocabulary Speech
  Recognition},'' \emph{arXiv preprint arXiv::1804.03209}, Apr. 2018.

\bibitem{voxceleb}
A.~Nagrani, J.~S. Chung, and A.~Zisserman, ``Voxceleb: A large-scale speaker
  identification dataset,'' in \emph{Interspeech}, 2017, pp. 2616--2620.

\bibitem{gt2013gtzan}
G.~Tzanetakis and P.~Cook, ``Musical genre classification of audio signals,''
  \emph{IEEE Speech Audio Process.}, vol.~10, no.~5, 2002.

\bibitem{nsynth2017}
J.~Engel, C.~Resnick, A.~Roberts, S.~Dieleman, M.~Norouzi, D.~Eck, and
  K.~Simonyan, ``Neural audio synthesis of musical notes with {W}ave{N}et
  autoencoders,'' in \emph{ICML}, 2017, pp. 1068--1077.

\bibitem{alain2016understanding}
G.~Alain and Y.~Bengio, ``Understanding intermediate layers using linear
  classifier probes,'' in \emph{ICLR}, 2017.

\bibitem{chenhui2019multilayers}
C.~Ma, X.~Mu, and D.~Sha, ``Multi-layers feature fusion of convolutional neural
  network for scene classification of remote sensing,'' \emph{IEEE Access},
  vol.~7, pp. 121\,685--121\,694, 2019.

\bibitem{weisen2021seen}
W.~Jiang, Y.~Zhang, and J.~T. Kwok, ``Seen: Few-shot classification with
  self-ensemble,'' in \emph{IJCNN}, 2021, pp. 1--8.

\bibitem{yue2018elastic}
Y.~Bai, S.~S. Bhattacharyya, A.~P. Happonen, and H.~Huttunen, ``Elastic neural
  networks: A scalable framework for embedded computer vision,'' in
  \emph{EUSIPCO}, Sep 2018.

\bibitem{Kumar2021DoSE}
A.~Kumar, Y.~Wang, V.~K. Ithapu, and C.~Fuegen, ``Do sound event
  representations generalize to other audio tasks? a case study in audio
  transfer learning,'' \emph{ArXiv}, vol. abs/2106.11335, 2021.

\bibitem{fonseca2021improving}
E.~Fonseca, A.~Ferraro, and X.~Serra, ``Improving sound event classification by
  increasing shift invariance in convolutional neural networks,'' \emph{arXiv
  preprint arXiv:2107.00623}, 2021.

\bibitem{chen20simclr}
T.~Chen, S.~Kornblith, M.~Norouzi, and G.~Hinton, ``A simple framework for
  contrastive learning of visual representations,'' in \emph{ICML}, vol. 119,
  13--18 Jul 2020, pp. 1597--1607.

\bibitem{he2020momentum}
K.~{He}, H.~{Fan}, Y.~{Wu}, S.~{Xie}, and R.~{Girshick}, ``Momentum contrast
  for unsupervised visual representation learning,'' in \emph{CVPR}, 2020, pp.
  9726--9735.

\bibitem{niizumi2021byol-a}
D.~Niizumi, D.~Takeuchi, Y.~Ohishi, N.~Harada, and K.~Kashino, ``Byol for
  audio: Self-supervised learning for general-purpose audio representation,''
  in \emph{IJCNN}, Jul 2021.

\bibitem{Liu2020Mockingjay}
A.~T. Liu, S.-w. Yang, P.-H. Chi, P.-c. Hsu, and H.-y. Lee, ``Mockingjay:
  Unsupervised speech representation learning with deep bidirectional
  transformer encoders,'' in \emph{ICASSP}, May 2020.

\bibitem{baevski2020wav2vec2}
A.~Baevski, Y.~Zhou, A.~Mohamed, and M.~Auli, ``wav2vec 2.0: {A} framework for
  self-supervised learning of speech representations,'' in \emph{NeurIPS},
  2020.

\bibitem{Hsu2021HuBERT}
W.-N. Hsu, B.~Bolte, Y.-H.~H. Tsai, K.~Lakhotia, R.~Salakhutdinov, and
  A.~Mohamed, ``Hubert: Self-supervised speech representation learning by
  masked prediction of hidden units,'' \emph{IEEE/ACM Trans. Audio, Speech,
  Language Process.}, p. 3451^^e2^^80^^933460, 2021.

\bibitem{spijkervet2021contrastive}
J.~Spijkervet and J.~A. Burgoyne, ``Contrastive learning of musical
  representations,'' \emph{arXiv preprint arXiv:2103.09410}, 2021.

\bibitem{youtube8m}
S.~Abu-El-Haija, N.~Kothari, J.~Lee, P.~Natsev, G.~Toderici, B.~Varadarajan,
  and S.~Vijayanarasimhan, ``Youtube-8m: A large-scale video classification
  benchmark,'' \emph{arXiv preprint arXiv:1609.08675}, 2016.

\bibitem{kim2019audiocaps}
C.~D. Kim, B.~Kim, H.~Lee, and G.~Kim, ``Audiocaps: Generating captions for
  audios in the wild,'' in \emph{NAACL-HLT}, 2019.

\bibitem{tolkova2021parsing}
I.~Tolkova, B.~Chu, M.~Hedman, S.~Kahl, and H.~Klinck, ``Parsing birdsong with
  deep audio embeddings,'' \emph{arXiv preprint arXiv:2108.09203}, 2021.

\bibitem{ravanelli2020paceplus}
M.~{Ravanelli}, J.~{Zhong}, S.~{Pascual}, P.~{Swietojanski}, J.~{Monteiro},
  J.~{Trmal}, and Y.~{Bengio}, ``Multi-task self-supervised learning for robust
  speech recognition,'' in \emph{ICASSP}, 2020, pp. 6989--6993.

\bibitem{bapna2021slam}
A.~Bapna, Y.~an~Chung, N.~Wu, A.~Gulati, Y.~Jia, J.~H. Clark, M.~Johnson,
  J.~Riesa, A.~Conneau, and Y.~Zhang, ``Slam: A unified encoder for speech and
  language modeling via speech-text joint pre-training,'' \emph{arXiv preprint
  arXiv:2110.10329}, 2021.

\bibitem{zhang2021bigssl}
Y.~Zhang, D.~S. Park, W.~Han, J.~Qin, A.~Gulati, J.~Shor, A.~Jansen, Y.~Xu,
  Y.~Huang, S.~Wang, Z.~Zhou, B.~Li, M.~Ma, W.~Chan, J.~Yu, Y.~Wang, L.~Cao,
  K.~C. Sim, B.~Ramabhadran, T.~N. Sainath, F.~Beaufays, Z.~Chen, Q.~V. Le,
  C.-C. Chiu, R.~Pang, and Y.~Wu, ``Bigssl: Exploring the frontier of
  large-scale semi-supervised learning for automatic speech recognition,''
  \emph{arXiv preprint arXiv:2109.13226}, 2021.

\bibitem{baevski2022data2vec}
A.~Baevski, W.-N. Hsu, Q.~Xu, A.~Babu, J.~Gu, and M.~Auli, ``data2vec: A
  general framework for self-supervised learning in speech, vision and
  language,'' \emph{arXiv preprint arXiv:2202.03555}, 2022.

\bibitem{wang2022universal}
L.~Wang, P.~Luc, Y.~Wu, A.~Recasens, L.~Smaira, A.~Brock, A.~Jaegle, J.-B.
  Alayrac, S.~Dieleman, J.~Carreira, and A.~van~den Oord, ``Towards learning
  universal audio representations,'' in \emph{ICASSP}, 2022, pp. 4593--4597.

\bibitem{scheidwasserclow2021serab}
N.~Scheidwasser-Clow, M.~Kegler, P.~Beckmann, and M.~Cernak, ``Serab: A
  multi-lingual benchmark for speech emotion recognition,'' in \emph{ICASSP},
  2022, pp. 7697--7701.

\bibitem{Chen2021WavLM}
S.~Chen, C.~Wang, Z.~Chen, Y.~Wu, S.~Liu, Z.~Chen, J.~Li, N.~Kanda,
  T.~Yoshioka, X.~Xiao, J.~Wu, L.~Zhou, S.~Ren, Y.~Qian, Y.~Qian, J.~Wu,
  M.~Zeng, and F.~Wei, ``Wavlm: Large-scale self-supervised pre-training for
  full stack speech processing,'' \emph{arXiv preprint arXiv:2110.13900}, 2021.

\bibitem{yang2021superb}
S.~wen Yang, P.-H. Chi, Y.-S. Chuang, C.-I.~J. Lai, K.~Lakhotia, Y.~Y. Lin,
  A.~T. Liu, J.~Shi, X.~Chang, G.-T. Lin, T.-H. Huang, W.-C. Tseng, K.~tik Lee,
  D.-R. Liu, Z.~Huang, S.~Dong, S.-W. Li, S.~Watanabe, A.~Mohamed, and
  H.~yi~Lee, ``{SUPERB: Speech Processing Universal PERformance Benchmark},''
  in \emph{Interspeech}, 2021, pp. 1194--1198.

\bibitem{oord2018cpc}
A.~van~den Oord, Y.~Li, and O.~Vinyals, ``Representation learning with
  contrastive predictive coding,'' \emph{arXiv preprint arXiv:1807.03748},
  2018.

\bibitem{turian2022hear}
J.~Turian, J.~Shier, H.~R. Khan, B.~Raj, B.~W. Schuller, C.~J. Steinmetz,
  C.~Malloy, G.~Tzanetakis, G.~Velarde, K.~McNally, M.~Henry, N.~Pinto,
  C.~Noufi, C.~Clough, D.~Herremans, E.~Fonseca, J.~Engel, J.~Salamon,
  P.~Esling, P.~Manocha, S.~Watanabe, Z.~Jin, and Y.~Bisk, ``Hear: Holistic
  evaluation of audio representations,'' \emph{arXiv preprint
  arXiv:2203.03022}, 2022.

\bibitem{ford2019deepresidual}
L.~Ford, H.~Tang, F.~Grondin, and J.~R. Glass, ``A deep residual network for
  large-scale acoustic scene analysis,'' in \emph{Interspeech}.\hskip 1em plus
  0.5em minus 0.4em\relax {ISCA}, 2019, pp. 2568--2572.

\bibitem{zhang2018mixup}
H.~Zhang, M.~Cisse, Y.~N. Dauphin, and D.~Lopez-Paz, ``mixup: Beyond empirical
  risk minimization,'' in \emph{ICLR}, 2018.

\bibitem{tokozume2018between}
Y.~Tokozume, Y.~Ushiku, and T.~Harada, ``Between-class learning for image
  classification,'' in \emph{CVPR}, Jun 2018.

\bibitem{logsumexp}
G.~C. Calafiore and L.~El~Ghaoui, \emph{Optimization Models}.\hskip 1em plus
  0.5em minus 0.4em\relax Cambridge University Press, 2014.

\bibitem{koizumi2020t6ntt}
\BIBentryALTinterwordspacing
Y.~Koizumi, D.~Takeuchi, Y.~Ohishi, N.~Harada, and K.~Kashino, ``The {NTT}
  {DCASE2020} challenge task 6 system: Automated audio captioning with keywords
  and sentence length estimation,'' DCASE2020 Challenge, Tech. Rep., 2020.
  [Online]. Available: \url{https://arxiv.org/abs/2007.00225}
\BIBentrySTDinterwordspacing

\bibitem{koutini2021receptive}
K.~Koutini, H.~Eghbal-zadeh, and G.~Widmer, ``Receptive field regularization
  techniques for audio classification and tagging with deep convolutional
  neural networks,'' \emph{IEEE/ACM Trans. Audio, Speech, Language Process.},
  vol.~29, pp. 1987--2000, 2021.

\bibitem{turian2021torchsynth}
J.~Turian, J.~Shier, G.~Tzanetakis, K.~McNally, and M.~Henry, ``One billion
  audio sounds from {GPU}-enabled modular synthesis,'' in \emph{DAFx2020}, Sep.
  2021.

\bibitem{voxforge}
\BIBentryALTinterwordspacing
K.~MacLean, ``Voxforge,'' 2018. [Online]. Available:
  \url{http://www.voxforge.org/home}
\BIBentrySTDinterwordspacing

\bibitem{cao2014cremad}
H.~Cao, D.~G. Cooper, M.~K. Keutmann, R.~C. Gur, A.~Nenkova, and R.~Verma,
  ``Crema-d: Crowd-sourced emotional multimodal actors dataset,'' \emph{IEEE
  Trans. Affective Comput.}, vol.~5, no.~4, pp. 377--390, 2014.

\bibitem{kereliuk2015music}
C.~Kereliuk, B.~L. Sturm, and J.~Larsen, ``Deep learning and music
  adversaries,'' \emph{IEEE Trans. Multimedia}, vol.~17, no.~11, p.
  2059^^e2^^80^^932071, Nov 2015.

\bibitem{sturm2013gtzansplit}
B.~L. Sturm, ``The gtzan dataset: Its contents, its faults, their effects on
  evaluation, and its future use,'' \emph{ArXiv}, vol. abs/1306.1461, 2013.

\bibitem{akiba2019optuna}
T.~Akiba, S.~Sano, T.~Yanase, T.~Ohta, and M.~Koyama, ``Optuna: A
  next-generation hyperparameter optimization framework,'' in \emph{SIGKDD},
  2019.

\bibitem{ViT}
A.~Dosovitskiy, L.~Beyer, A.~Kolesnikov, D.~Weissenborn, X.~Zhai,
  T.~Unterthiner, M.~Dehghani, M.~Minderer, G.~Heigold, S.~Gelly, J.~Uszkoreit,
  and N.~Houlsby, ``An image is worth 16x16 words: Transformers for image
  recognition at scale,'' in \emph{ICLR}, 2021.

\bibitem{DeiT}
H.~Touvron, M.~Cord, M.~Douze, F.~Massa, A.~Sablayrolles, and H.~J{\'{e}}gou,
  ``Training data-efficient image transformers {\&} distillation through
  attention,'' in \emph{ICML}, vol. 139, 2021, pp. 10\,347--10\,357.

\bibitem{Arandjelovic2017l3net}
R.~Arandjelovic and A.~Zisserman, ``Look, listen and learn,'' in \emph{ICCV},
  Oct 2017.

\bibitem{freesound}
F.~Font, G.~Roma, and X.~Serra, ``Freesound technical demo,'' in \emph{ACM-MM},
  2013, p. 411^^e2^^80^^93412.

\bibitem{Panayotov2015LibrispeechAA}
V.~Panayotov, G.~Chen, D.~Povey, and S.~Khudanpur, ``Librispeech: An asr corpus
  based on public domain audio books,'' in \emph{ICASSP}, 2015, pp. 5206--5210.

\end{thebibliography}


\vspace{-10pt}
\begin{IEEEbiography}[{\includegraphics[bb=0 0 640 706,width=1in,height=1.25in,clip,keepaspectratio]{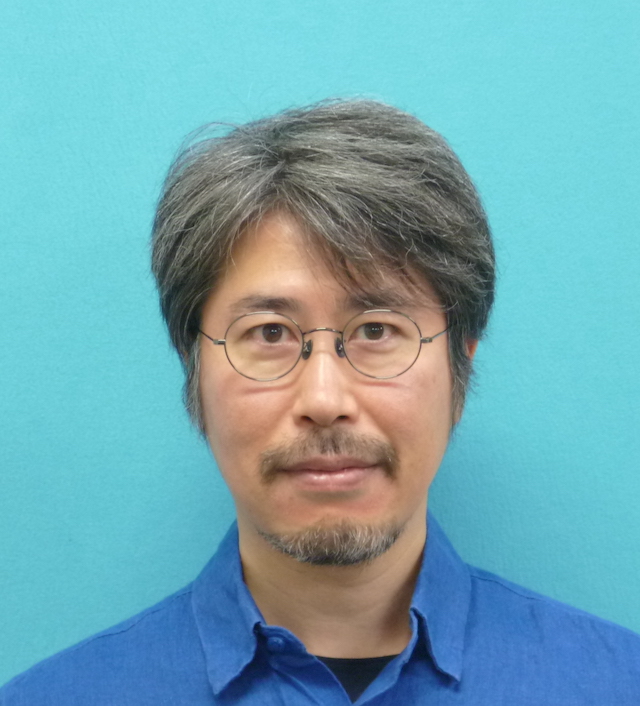}}]{Daisuke Niizumi}
Daisuke Niizumi received the B.S., and M.S., degrees from the Department of Computer Science and Systems Engineering of the Kyushu Institute of Technology in 1995 and 1997, respectively. From 1997 to 2020, he was a senior software and machine learning engineer/manager in several consumer electronics companies. He joined NTT Corporation in 2020. His research interests include representation learning, multimodal deep learning, and anomaly detection.
\end{IEEEbiography}

\vspace{-10pt}
\begin{IEEEbiography}[{\includegraphics[bb=0 0 65 83,width=1in,height=1.25in,clip,keepaspectratio]{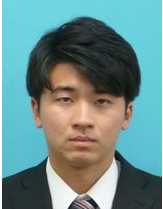}}]{Daiki Takeuchi}
Daiki Takeuchi received the B.S., and M.S., degrees from the Department of Intermedia Art and Science of Waseda University in 2018 and 2020, respectively.  He joined NTT Corporation in 2020. His research interests include signal processing and machine learning for multimodal information processing related to audio.
\end{IEEEbiography}

\vspace{-10pt}
\begin{IEEEbiography}[{\includegraphics[bb=0 0 602 741,width=1in,height=1.25in,clip,keepaspectratio]{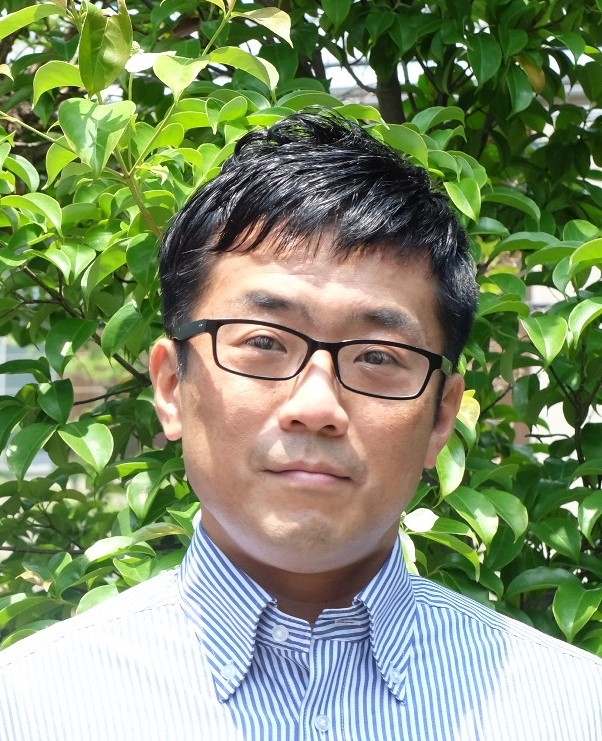}}]{Yasunori Ohishi}
Yasunori Ohishi (M '12) received his Ph.D. degree from Nagoya University in 2009. Since joining NTT in 2009, he has been researching speech and audio signal processing. His research interests generally concern audio event detection, music information retrieval, and crossmodal learning with audio applications. 
\end{IEEEbiography}

\vspace{-10pt}
\begin{IEEEbiography}[{\includegraphics[bb=0 0 224 289,width=1in,height=1.25in,clip,keepaspectratio]{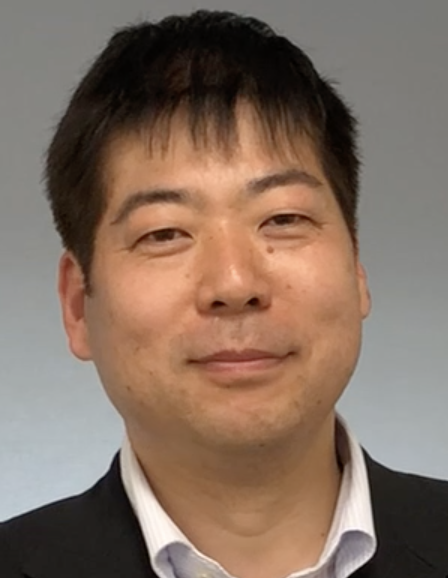}}]{Noboru Harada} Noboru Harada (SM '18) received the B.S. and M.S. degrees in computer science and systems engineering from the Kyushu Institute of Technology, Fukuoka, Japan, in 1995 and 1997, respectively, and the Ph.D. degree in computer science from the University of Tsukuba, Ibaraki, Japan, in 2017. Since joining NTT Corporation, Tokyo, Japan, in 1997, he has been involved with research on speech and audio signal processing, such as high-efficiency coding, lossless compression, and acoustic event detection, including anomaly sound detection.
\end{IEEEbiography}

\vspace{-10pt}
\begin{IEEEbiography}[{\includegraphics[bb=0 0 1590 1909,width=1in,height=1.25in,clip,keepaspectratio]{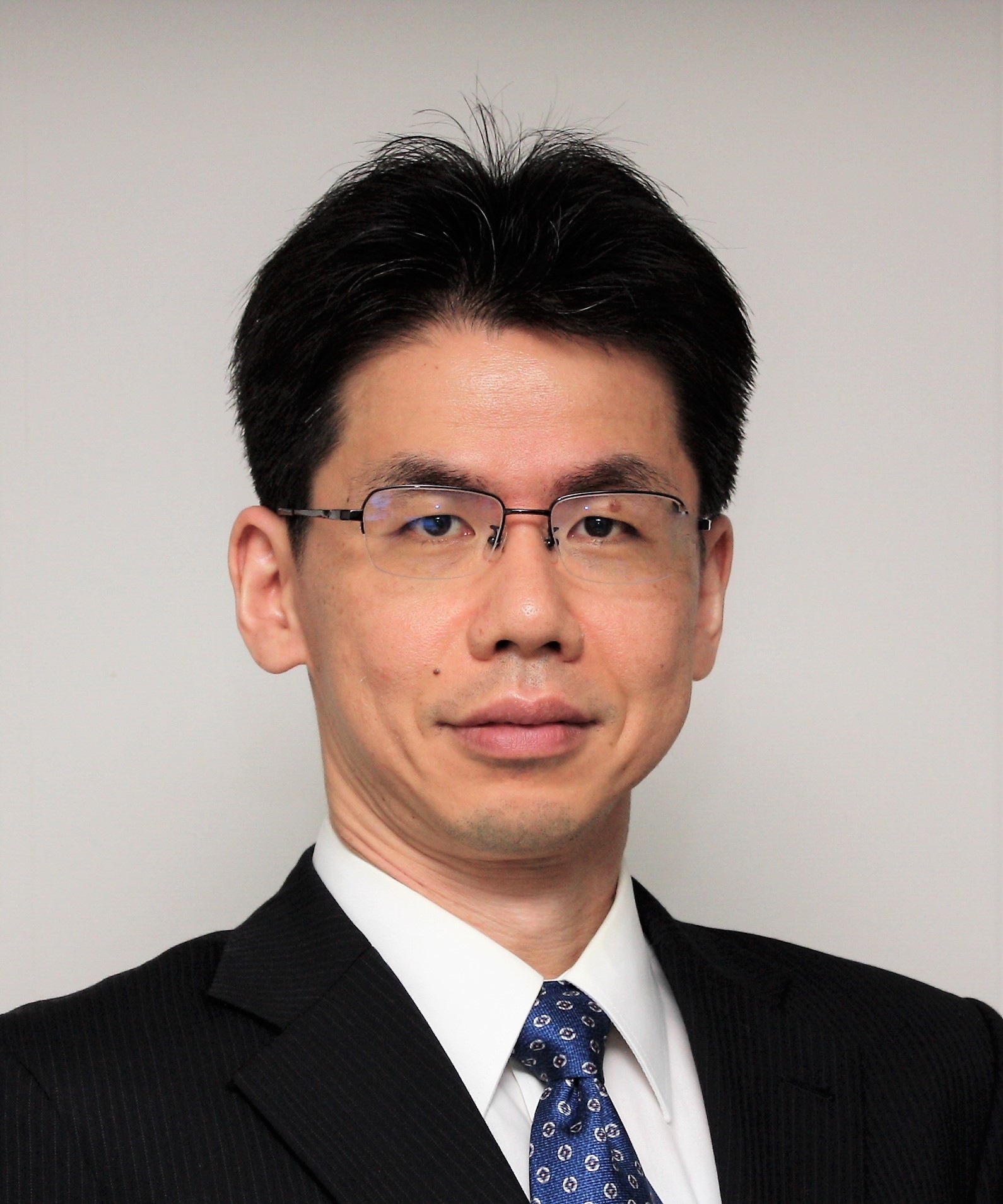}}]{Kunio Kashino}
Kunio Kashino (SM '05) received his Ph.D. degree from the University of Tokyo in 1995. He is currently a Senior Distinguished Researcher at Nippon Telegraph and Telephone Corporation and a visiting professor at National Institute of Informatics, Japan. His research interests include audio and crossmodal information processing, media search, and biomedical informatics. He is a member of Association for Computing Machinery (ACM), and a Fellow of the Institute of Electronics, Information and Communication Engineers (IEICE). He received IEEE Transactions on Multimedia Paper Award in 2004, Maejima Award in 2010, Kenjiro Takayanagi Achivement Award in 2016, IEICE Achievement Award in 2002 and 2017, and the Commendation for Science and Technology by the Minister of Education, Culture, Sports, Science and Technology in 2019.
\end{IEEEbiography}

\vfill

\end{document}